\documentclass[sigconf]{acmart}
\AtBeginDocument{%
  }

\copyrightyear{2026}
\acmYear{2026}
\setcopyright{cc}
\setcctype{by}
\acmConference[CHI '26]{Proceedings of the 2026 CHI Conference on Human Factors in Computing Systems}{April 13--17, 2026}{Barcelona, Spain}
\acmBooktitle{Proceedings of the 2026 CHI Conference on Human Factors in Computing Systems (CHI '26), April 13--17, 2026, Barcelona, Spain}
\acmDOI{10.1145/3772318.3791579}
\acmISBN{979-8-4007-2278-3/2026/04}

\usepackage{xspace}
\usepackage{multirow}

\newcommand{\systemname}{VizCrit\xspace}
\newcommand{\directed}{solution-centered\xspace}
\newcommand{\neutral}{awareness-centered\xspace}
\newcommand{\Directed}{Solution-centered\xspace}
\newcommand{\Neutral}{Awareness-centered\xspace}
\newcommand{\neutraldirected}{awareness- and solution-centered\xspace}
\newcommand{\baseline}{textbook-based\xspace}

\usepackage{xcolor}

\newif\ifsubmit
\submittrue
\ifsubmit
    \newcommand{\goal}[1]{}
    \newcommand{\todo}[1]{}
    \newcommand{\jane}[1]{}
    \newcommand{\fig}[1]{}
    \newcommand{\mingyi}[1]{}
    \newcommand{\steven}[1]{}
    \newcommand{\hazel}[1]{}
    \newcommand{\maitraye}[1]{}
    \newcommand{\revised}[1]{#1}
    \newcommand{\revi}[1]{#1}
\else
    \newcommand{\goal}[1]{\textbf{\textcolor{ACMDarkBlue}{goal: #1}}}
    \newcommand{\todo}[1]{\textbf{\textcolor{ACMDarkBlue}{TODO: #1}}}
    \newcommand{\fig}[1]{\textbf{\textcolor{orange}{figure: #1}}}
    \newcommand{\jane}[1]{\textbf{\textcolor{violet}{jane: #1}}}
    \newcommand{\mingyi}[1]{\textbf{\textcolor{cyan}{mingyi: #1}}}
    \newcommand{\steven}[1]{\textbf{\textcolor{green}{steven: #1}}}
    \newcommand{\hazel}[1]{\textbf{\textcolor{pink}{hazel: #1}}}
    \newcommand{\maitraye}[1]{\textbf{\textcolor{lime}{maitraye: #1}}}
    \newcommand{\revised}[1]{\textcolor{blue}{#1}}
    \newcommand{\revi}[1]{\textcolor{teal}{#1}}
\fi

\newif\ifred
\redfalse
\ifred
    \newcommand{\revise}[1]{\textcolor{ACMRed}{#1}}
\else
    \definecolor{red}{HTML}{000000}
    \newcommand{\revise}[1]{#1}
\fi

\begin{document}

\title{VizCrit: Exploring Strategies for Displaying Computational Feedback in a Visual Design Tool}


\author{Mingyi Li}
\orcid{0000-0002-7192-7407}
\affiliation{%
  \institution{Northeastern University}
  \city{Boston}
  \state{Massachusetts}
  \country{USA}
}
\email{li.mingyi2@northeastern.edu}

\author{Mengyi Chen}
\orcid{0009-0006-0133-6384} 
\affiliation{%
  \institution{University of Pennsylvania}
  \city{Philadelphia}
  \state{Pennsylvania}
  \country{USA}
}
\email{mengyic@seas.upenn.edu}

\author{Sarah Luo}
\orcid{0009-0004-3536-5429}
\affiliation{%
  \institution{Purdue University}
  \city{West Lafayette}
  \state{Indiana}
  \country{USA}
}
\email{luo459@purdue.edu}

\author{Yining Cao}
\orcid{0000-0002-3962-2830}
\affiliation{%
  \institution{University of California, San Diego}
  \city{La Jolla}
  \state{California}
  \country{USA}
}
\email{rimacyn@ucsd.edu}

\author{Haijun Xia}
\orcid{0000-0002-9425-0881}
\affiliation{%
  \institution{University of California, San Diego}
  \city{La Jolla}
  \state{California}
  \country{USA}
}
\email{haijunxia@ucsd.edu}

\author{Maitraye Das}
\orcid{0000-0002-8640-0544}
\affiliation{%
  \institution{Northeastern University}
  \city{Boston}
  \state{Massachusetts}
  \country{USA}
}
\email{ma.das@northeastern.edu}

\author{Steven P. Dow}
\orcid{0000-0002-1354-9866}
\affiliation{%
  \institution{University of California, San Diego}
  \city{La Jolla}
  \state{California}
  \country{USA}
}
\email{spdow@ucsd.edu}

\author{Jane L. E}
\orcid{0000-0001-6868-7732}
\affiliation{%
  \institution{National University of Singapore}
  \city{Singapore}
  \country{Singapore}
}
\email{ejane@nus.edu.sg}

\renewcommand{\shortauthors}{Li et al.}

\begin{abstract}
Visual design instructors often provide multi-modal feedback, mixing annotations with text. Prior theory emphasizes the importance of actionable feedback, where ``actionability'' lies on a spectrum---from surfacing relevant design concepts 
to suggesting concrete fixes. 
\revi{How might creativity tools implement annotations that support such feedback, and how does the actionability of feedback impact novices' process-related behaviors, perceptions of
creativity, learning of design principles, and overall outcomes?} We introduce \textit{\systemname}, a system for providing computational feedback that supports the actionability spectrum, realized through algorithmic issue detection and visual annotation generation. 
In a between-subjects study (N=36), novices revised a design under one of three conditions: \baseline, \neutral, or \directed feedback.
We found that \directed feedback \revised{led to fewer design issues and higher self-perceived creativity compared with \baseline feedback, although expert ratings on creativity showed no significant differences.}  
We discuss the implications for AI in Creativity Support Tools, \revi{including the potential of calibrating feedback actionability 
to help novices balance productivity with learning, growth, and developing design awareness.}
\end{abstract}



\begin{teaserfigure}
  \centering
  \includegraphics[width=\textwidth]{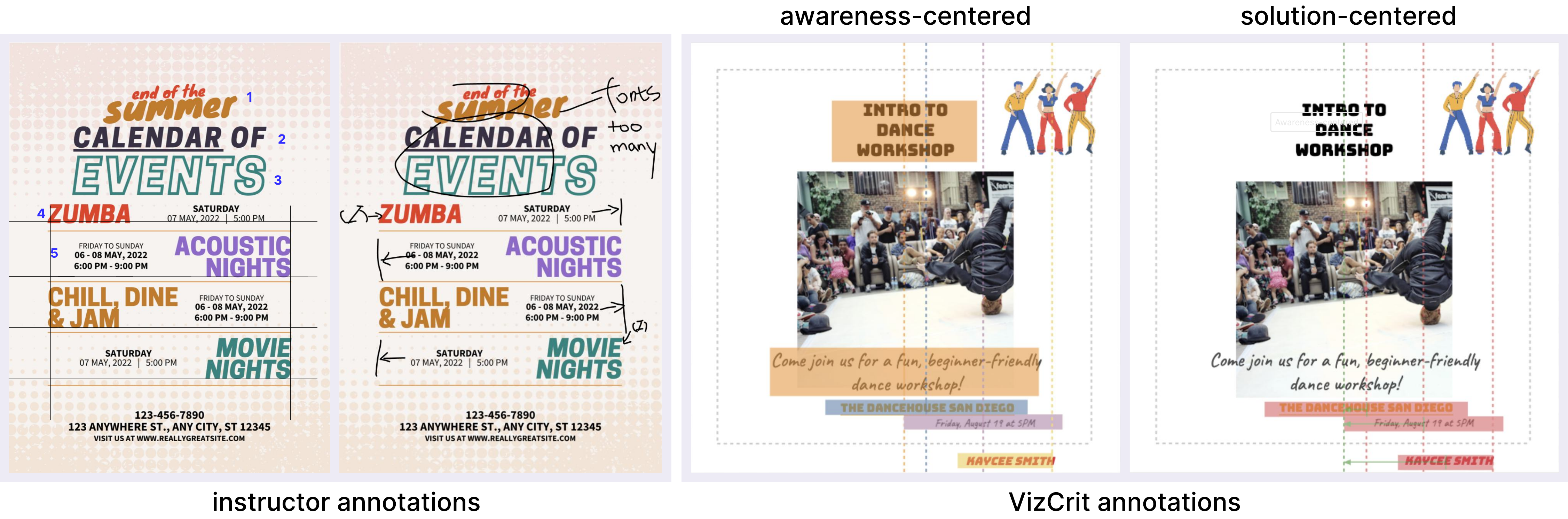}
  \caption{We introduce \textit{VizCrit}, a design feedback tool that offers feedback with three levels of actionability: \textit{\baseline} feedback as static text, and \textit{\neutral} and \textit{\directed} feedback as adaptive visual annotations. 
  To design the interactive annotations, 
  we observed how expert designers and instructors provide situated feedback (left), then collaboratively co-designed a set of visual annotations  (right) for four core design principles (Alignment is shown here). For each annotation, we designed algorithms for heuristically computing the annotations to be displayed as overlays on the visual design (including issue detection for \directed feedback). \revi{The user evaluation with novices explores how different \textit{actionability} in feedback influences novices' process-related behaviors, learning of principles, perceptions of creativity, and overall outcomes.}  
  }
  \Description{This teaser figure demonstrates VizCrit, a design feedback tool that provides textbook-like, awareness-centered, and solution-centered feedback. There are two instructor annotations on the left, which contain their hand-written sketches on designs. On the left of the two, there are straight lines dividing the canvas into different regions and numbers marking different sections of the design. This is an example of awareness-centered alignment feedback, and VizCrit approaches it by highlighting the text boxes in different colors for different alignment groups. On the right of the two instructor's annotations, the design now shows circles, arrows asking texts to be aligned, vertical lines indicating the alignment axes, and texts "fonts too many." This is an example of solution-centered feedback, and VizCrit approaches it by highlighting the problematic texts in red, with green arrows pointing to the suggested alignment axis marked with dashed lines.}
  \label{fig:teaser}
\end{teaserfigure}

\maketitle

\section{Introduction}\label{sec:intro}
In design education, effective feedback is central to scaffolding creative growth~\cite{schon1983reflective,barrett1988comparison,dannels2008critiquing}.
Instructor- and peer-led studio critiques (or ``crits'') can help students reflect, iterate, and develop disciplinary judgment, both during the act of designing and after drafts are complete~\cite{schon1983reflective,barrett1988comparison,dannels2008critiquing}. Instructors often provide multi-modal feedback directly on students' work, externalized as a mix of visual annotations~\cite{suwa2002external} and spoken or written commentary to highlight principles in context. These practices do more than convey corrections: they shape how novices see their work, explore alternatives, and build evaluative judgment~\cite{dannels2008beyond,gray2013informal}.  
\revi{To investigate the potential for computational scaffolding, we explore how multi-modal feedback, including visual annotations, can be integrated into a design tool and how this impacts design learners.}

\revi{Existing research characterizes high-quality feedback as} \emph{specific}, \emph{justified}, and \emph{actionable} to effectively guide learning~\cite{nicol2006formative,hattie2007power,shute2008focus,sadler1989feedback}.
\revi{Given recent advances in AI, feedback of all styles is becoming readily available. We therefore explore} \emph{actionability}---how directly feedback prescribes changes---because it can shape both the creative outcomes and the potential for learning. 
On one hand, instructors typically 
avoid imposing specific solutions on students; they ask students to notice issues and reason about how design choices relate to underlying concepts~\cite{dannels2008beyond,dannels2008critiquing}. 
On the other hand, AI tools could be designed to provide a range of feedback, with a norm leaning towards prescriptive feedback with concrete fixes that can accelerate practice
~\cite{schwartz2004inventing,schwartz2011practices}, but 
\revi{might reduce opportunities for} reflection ~\cite{jansson1991design,purcell1996design, nicol2006formative,winstone2017itd}.

While prior work in design education has offered hints on potential tradeoffs between 
different levels of feedback actionability~\cite{sadler1989feedback,shute2008focus,dannels2008beyond,nelson2009nature,wu2020feedback}, we still lack \revi{solid empirical data on how different choices} 
affect novices' learning and creativity.
We explore key questions within the context of a visual design tool: 
\begin{itemize}
    \item \textbf{[RQ1]} 
    How might we design visual annotations to support different levels of actionability? and,
    \item \textbf{[RQ2]} How does the actionability of computational feedback affect \revi{design process-related behaviors, perceptions of creativity, learning of principles, and overall outcomes?}
\end{itemize}
As tools like generative AI can now provide immediate, always-available, and the full spectrum of design feedback ~\cite{fischer1994turning, davis2016drawing}, \revi{these concerns about different feedback styles have important implications for how to position AI within design workflows and human-AI collaboration~\cite{gero2022sparks,biermann2022tool, tone2025integrating}.}  
Understanding the impacts of different feedback actionability \revi{can help Creativity Support Tools (CSTs) designers 
tailor feedback} in ways that support both productivity and creative growth~\cite{li2023beyond,bennett2024painting}. 

To answer \textbf{[RQ1]}, 
we \revi{observed how visual design experts use visual annotations as part of their critique process}~\cite{dannels2008beyond,dannels2008critiquing}.
We conducted co-design interviews with design experts (N=9) who provided critiques using different feedback actionability on a set of designs.
Our analysis of these critiques informed the designs of our visual annotations for hierarchy, alignment, whitespace, and unity design principles. We embed these into \textit{VizCrit}, a prototype \revi{for a visual design tool that delivers on-demand and principled feedback. \textit{VizCrit} provides feedback that spans} the actionability spectrum:
static, \textit{\baseline} feedback via text only, and adaptive visual annotations that are either \textit{\neutral}, focusing on drawing awareness to design principles, or \textit{\directed}, both calling out issues within a design and suggesting a concrete change.
To implement these annotations, we carefully designed algorithms to computationally generate the expert-informed annotations, and for the \directed annotations, to detect issues within a visual design, and to propose reasonable fixes.

For \textbf{[RQ2]}, drawing on prior work in education and design, we hypothesize a tradeoff: \directed feedback would improve immediate performance, confidence, and polish~\cite{nelson2009nature,wu2020feedback,shen2021teacher,e2024timing}, but perhaps at the cost of learning, creativity, and ownership~\cite{jansson1991design,purcell1996design}. In contrast, \neutral feedback would be harder to act on in the moment but could scaffold better reflection, transfer, and the development of evaluative judgment~\cite{sadler1989feedback,shute2008focus}. We also hypothesize that generic \baseline feedback 
would leave too big of a gap for novices to close, or to successfully transfer what they learn to address design principle violations~\cite{black1998assessment,hattie2007power}.
\revi{Using \textit{VizCrit} allows us to move beyond theoretical accounts to collect empirical data on how feedback actionability impacts novices' design behaviors.}
To investigate these hypotheses, we conducted a between-subjects study with novice designers (N=36) using \textit{\systemname} to revise a flawed visual design. Participants were randomly assigned to one of three conditions: \baseline, \neutral, or \directed feedback\revise{, and were asked to revise a design using the feedback provided.}
To understand the effects of feedback actionability, we measured participants' \revi{process-related behaviors (e.g.,} interaction timing and frequency of the feedback requests), self- and expert-rated creativity, design principle learning \revi{from knowledge tests}, and design outcomes \revi{(e.g., number of issues and degree of change)}.

\revised{We found that \directed participants gave significantly higher ratings on self-assessed creativity than \baseline participants, although there were no statistically significant differences in how independent experts assessed creativity across conditions. Our behavioral data on how participants responded to feedback} suggests that \directed participants may have developed a false sense of creativity, stemming from their trust in the system's suggestions \revi{and feelings of progress}.
Inline with our hypotheses, \directed participants showed better design outcomes \revised{as measured by a statistically significant decrease in design issues compared to \baseline participants.} 
Contrary to our hypotheses that \directed feedback would hinder learning, \revised{there were no significant differences in the learning gains across conditions. Our interview data suggests that 
\directed participants still reflected on how they applied principles to their own designs.}

We discuss implications for leveraging AI within CSTs and how to strike a balance between providing immediate assistance and helping novices in the long term to develop a critical eye and deepen their creative expertise.

This paper makes the following contributions:  
\begin{itemize}
  \item \textbf{Co-design study of feedback annotations with experts:} 
  \revi{We uncover insights from visual design experts and} contribute a set of visual annotations across \baseline, \neutraldirected feedback actionability.
  \item \textbf{System design and implementation:} We introduce \textit{\systemname}, a prototype that delivers feedback as visual annotations \revi{with linked text}, supported by algorithms for detecting issues and generating suggestions.
  \item \textbf{Study on feedback actionability with design novices:} 
  In a between-subjects study (N=36), 
    \revised{we found preliminary evidence, despite our hypotheses, that \directed feedback led to fewer design issues and increased self-perceived creativity compared with \baseline feedback. However, expert ratings on creativity showed no significant differences}, 
    \revi{suggesting that more actionability may inflate novices' perceived creativity through feedback-driven progress.} 
\end{itemize}

\section{Related Work}\label{sec:related}

We review related literature on feedback in education as well as design, highlighting the dimensions of ``actionability'' in each domain. We then review work exploring the potential of computational guidance for creativity, \revise{generative AI for design feedback}, and the potential influence of such tools on creativity and learning.  

\subsection{The ``Actionability'' Factor in Design Feedback}

Educational research emphasizes the centrality of feedback for learning, yet its effects vary with content, timing, and learner expertise~\cite{sadler1989feedback, cutumisu2018impact, black1998assessment, hattie2007power}. Foundational work emphasizes that feedback only improves learning if learners can close the gap between current and desired performance~\cite{sadler1989feedback,black1998assessment,hattie2007power}. 
\revi{Educational researcher Shute ~\cite{shute2008focus} defines different feedback styles, from directive (tells learners what to fix) to facilitative (prompts learners to engage in their own correction). In our work, we expand on this to define a spectrum of actionability in feedback from the most generic, unactionable (i.e. like a textbook) to \neutraldirected feedback and beyond.}
Studies show that while facilitative feedback such as principle-based comments support reflection, students revise more effectively when feedback includes concrete suggestions~\cite{nelson2009nature,wu2020feedback}; in STEAM classrooms, directive feedback also produced higher creative gains than opinion-only feedback~\cite{shen2021teacher}. However,
directive feedback reduced students' autonomy and self-reflection on the learning process, as the main focus became fixing errors \cite{williams2024}.
These frameworks show that while directive input can accelerate revision, overly prescriptive feedback risks limiting learners' engagement in reflection and transfer.


Critique structures help learners reflect and iterate ~\cite{mcdonald2022uncertain, barrett1988comparison, lousberg2020reflection, schon1983reflective}.
In line with education, work in design highlights a continuum from comments that raise awareness of principles to those that propose specific changes. 
Researchers found that design studio critique often emphasizes principle-based comments that build disciplinary understanding without prescribing exact moves~\cite{dannels2008beyond, dannels2008critiquing}. \citet{nguyen2017fruitful} similarly showed that open-ended or affective framing of critique can foster receptivity and preserve a sense of ownership in student work. 
At the same time, studies of fixation warn that prescriptive examples can constrain creativity: designers often remain anchored to given examples, highlighting how prescriptive examples can anchor design trajectories in unproductive ways~\cite{jansson1991design,purcell1996design}.
\revi{Motivated by prior work, we compare the tradeoffs in learning, creativity, and performance in the visual design domain, particularly when facilitated by computational support.} 

\subsection{Computational Guidance in Creativity Support Tools}
By embedding certain assumptions and abstractions, creative support tools hold power over how users create \cite{li2023beyond, winner1980artifacts}. Therefore, they influence not only outcomes but also creative ownership and agency. HCI has long argued that tools should empower users, not merely optimize productivity \cite{cherry14CSI,frich2019mapping,constanzachock2020design}.
The growth in AI-driven creativity support intensifies these concerns: when systems readily provide ready-made solutions or feedback, they can accelerate work but risk shifting authorship away from the creator~\cite{bennett2024painting,biermann2022tool,gero2022sparks}.

\revi{Considering the lens of power in CSTs ~\cite{li2023beyond} provides insights on feedback actionability and how it imposes different ways of thinking.}
Existing tools cover a spectrum from \emph{surfacing awareness} 
\revi{(which gives users power over their next move)} to guidance that \emph{proposes potential answers} 
\revi{(where the feedback imposes power over the user's next move)}.
Awareness-centered tools aim to widen perception and scaffold reflection: e.g., adaptive conceptual prompts in \textit{Sh{\"o}wn}~\cite{ngoon2021shown}, principle-oriented critics and analytic overlays~\cite{fischer1991critics,lee2020guicomp}, and composition guidance for photography~\cite{e2020adaptivephotocomposition}. Other systems offer more specific suggestions to accelerate revision, such as interactive layout suggestions in \textit{DesignScape}~\cite{odonovan2015designscape} or guidance towards achieving specific portrait lighting styles~\cite{e2019portraitlighting}. 

\revise{Recently, a growing body of work investigates generative AI for feedback in various contexts, such as UI and web design \cite{duan24UICrit, duan24uifeedback, wu24UIClip, huh24designChecker}, 3D design~\cite{chen24MemoVis}, presentation slides \cite{zhang25slideAudit}, and art education \cite{zheng25artwork}.
Most related to our work, \textit{UICrit} \cite{duan24UICrit} generates design critiques for UI screenshots by collecting a dataset of critiques and using few-shot and visual prompting techniques with LLMs. Also related, \textit{DesignChecker} \cite{huh24designChecker} provides suggestions on web design issues by applying design heuristics-based algorithms. 
Most of these focus on a few specific feedback designs (often \directed) and tend to be primarily text-based. Our research builds on this prior work by studying the difference in impact across the spectrum of actionability and the potential of visual annotations for feedback.}

Since novices are still developing domain expertise and evaluative judgment~\cite{ericsson1993deliberate,goodwin1994professional}, they are especially susceptible to tool-related influence. Recent empirical work on feedback timing shows that when and how guidance appears (in-action versus on-action) risks novices' overreliance on the tool for creative decisions~\cite{e2024timing}.
These perspectives suggest that the \emph{form} of guidance matters for both learning and creativity.
Building on this literature, we use the awareness-versus-solution distinction as an analytical lens to examine \emph{how} guidance impacts novices' learning, creativity, and performance in visual design.
Inspired by work exploring how AI explanations can help calibrate trust and reduce overreliance~\cite{vasconcelos2023explanations}, we deliver guidance as \emph{visual annotations} that act as explanations, aiming to make system reasoning legible (mitigating overreliance) while preserving room for user judgment and exploration.

\subsection{The Role of Annotations in Design Communication}

Annotations are a key medium for design communication: by externalizing what matters in a layout, they help people see relations, surface issues, and coordinate critique. Work on sketching and external representations shows that marks ``on the page'' are not merely records but cognitive scaffolds that prompt noticing and reasoning about alternatives~\cite{suwa2002external,tversky2002sketches}. Annotations also carry a social–interpretive function: through practices like marking and coding, experts cultivate \emph{professional vision}, training novices to see work through disciplinary criteria and to connect abstract principles to concrete choices~\cite{goodwin1994professional}. 
In addition to this conceptual role, 
prior systems have used interactive annotations to make principles visible in context\revise{~\cite{e2020adaptivephotocomposition,lee2020guicomp,odonovan2015designscape,beaudouinlafon2023colorfield, ma25drawing}}. We build on this prior work by operationalizing annotations as a form of computational feedback that seeks to make design principles explicit while varying their actionability and studying how they impact novices' design processes and understanding of design principles.



\section{Co-Design Study on Feedback Annotations with Instructors}
\revi{
How might we design visual annotations to support
different levels of actionability?
To explore this question, we co-designed with experts} a set of annotations for design principles. For each principle, our goal was to create a parallel pair of annotation designs: 
one \neutral and one \directed. 
To ground these designs, we relied on experts in visual design for two stages: (1) collecting existing annotation examples from educational materials to generate preliminary design probes and (2) \revise{conducting a co-design study with experts where they annotated novice designs with} freeform, \neutral, and \directed feedback styles. 
All study materials can be found in Supplemental Materials.

\subsection{Design Probes} 
We first collected \revi{annotations that instructors and experts in design had already created around prior materials and then} performed an informal content analysis.
We searched for descriptions of design principles with examples using a range of keywords around design, principles, descriptions, visualizations, examples, etc. 
Across 43 articles, 10 books, and 7 videos, we extracted hundreds of examples and grouped them by principles (balance, alignment, whitespace, hierarchy, unity, readability). From these, we iteratively brainstormed a preliminary set of annotations. Balance was removed since many annotations from our content analysis mixed it with whitespace.
For each principle, we produced two awareness-centered and two solution-centered \revise{probes} to test in parallel, to reduce bias~\cite{dow2010parallel}.
\revise{
We used these artifacts as exploratory design probes to
elicit experts' annotation practices and their insights into the actionability of feedback.
The results were later synthesized into final design annotations for the \textit{\systemname} system.}

\begin{figure*}[t]
    \centering
    \includegraphics[width=\linewidth]{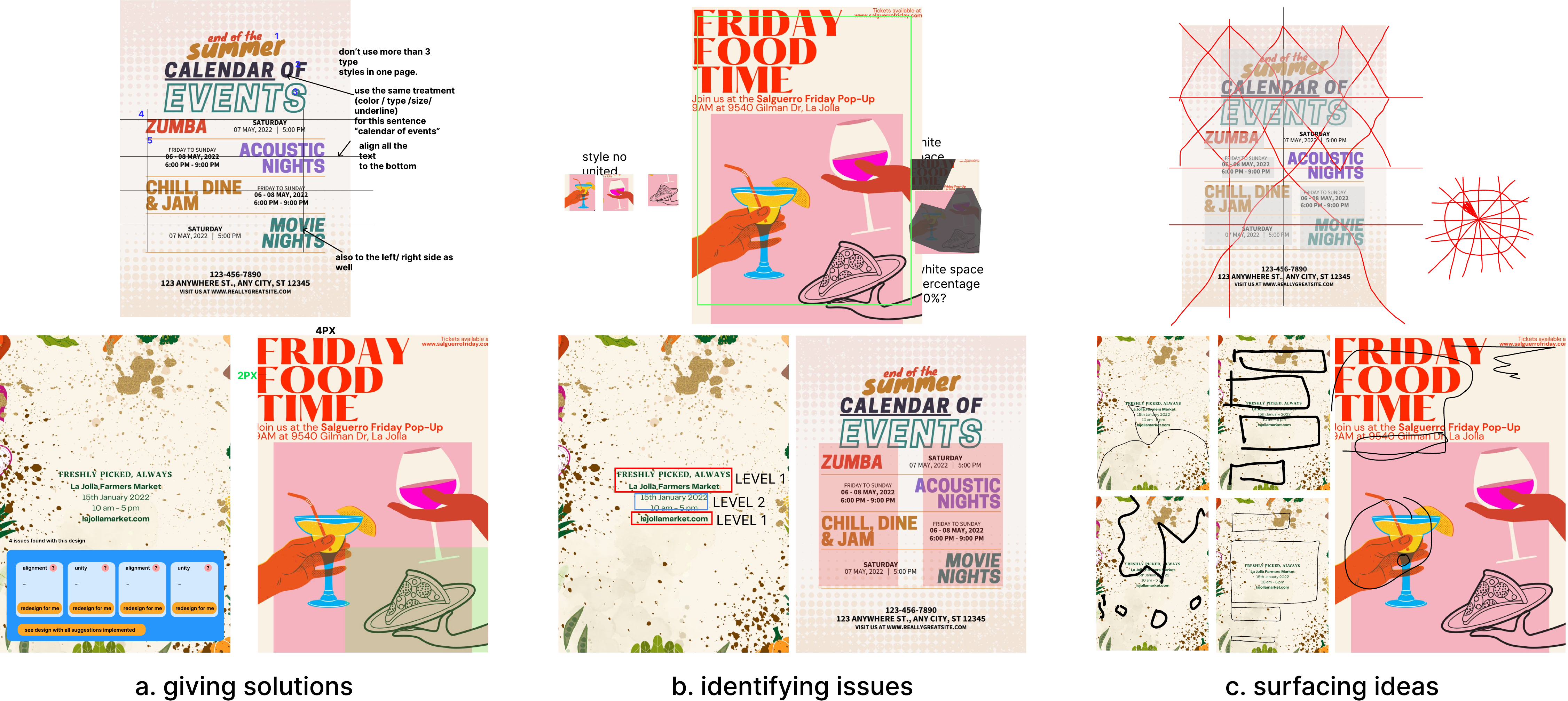}
    \caption{Expert instructors' freeform feedback shown here across the actionability spectrum: (a) \directed, (b) 
    calling awareness to issues, but not providing solutions
    , and (c) \neutral.
    }
    \Description{
    This figure demonstrates expert instructors' freeform feedback in three actionability types: a. giving solutions where instructors use lines, numbers, arrows, green boxes, or texts to show direct suggestions. b. Identifying issues where instructors use red rectangles, blue outlines, green rectangles, and texts to identify the design issues. c. surfacing ideas where instructors use grid structure, boxes, circles, lines to surface design ideas.
    }
    \label{fig:expertfreeform}
 \end{figure*}

\subsection{Participants}
To understand professional feedback practices and evaluate our annotation designs, we recruited 9 expert visual designers (6F, 3M; age: 21--50, $\mu = 28.0$) from online design communities or our personal networks.
Participants had an average of 6.4 years of experience with visual design and feedback-giving (range 1–20+), and included design instructors as well as professional product/UX/graphic designers.
We administered a pre-study screening survey to collect portfolios or websites, and then selected participants
based on their demonstrated expertise.

\subsection{Procedure}
Participants were first asked to review several novice designs and then provide feedback: first freeform, then \neutral, and finally \directed. This produced parallel examples across the spectrum and insights into design experts' feedback communication strategies.
We administered the co-design study on Figma, since it is widely used in the design community and supports both quick, low-fidelity annotations (e.g., freehand drawing) and high-fidelity operations (e.g., applying blur effects). 
When a change was difficult to execute quickly, the interviewer assisted by implementing it in Figma 
based on the participants' verbal descriptions 
and then confirmed its accuracy.

Each study was conducted over Zoom and lasted for approximately 1-1.5 hours. Each participant was compensated with a \$60 Amazon gift card. All sessions were screen-recorded to capture participants' actions and think-aloud protocols, which enabled us to observe participants' strategies in real time and collect rich qualitative data for analysis. 

\subsubsection{Interview}
After completing a demographics survey, participants were asked questions about their current process of providing feedback on visual designs. This included the integration of feedback into their visual design process or teaching, challenges they see novices encounter while applying or addressing feedback, and whether they used visual annotations or followed design principles to communicate feedback. 
 
\subsubsection{\revi{Freeform} Design Feedback}
Participants were asked to critique three novice designs in Figma, each with two principle violations (hierarchy, alignment, whitespace, unity, readability). 
Participants were asked to annotate or write their feedback for the novice designs using freeform feedback practices.
Researchers then explained the spectrum of \neutral to \directed feedback
and asked the participants to annotate those same designs using the \neutraldirected feedback styles.

\subsubsection{Feedback on \revi{Design Probes}}
Researchers showed and explained our preliminary annotation designs (two \neutral and two \directed per principle).
Each annotation idea was shown on a visual design that violates exactly that same principle. 
Participants then reviewed the designs and proposed their preferred annotations on a new canvas: either from scratch, or borrowing ideas from our design probes.

\begin{figure*}[th]
    \centering
    \includegraphics[width=\linewidth]{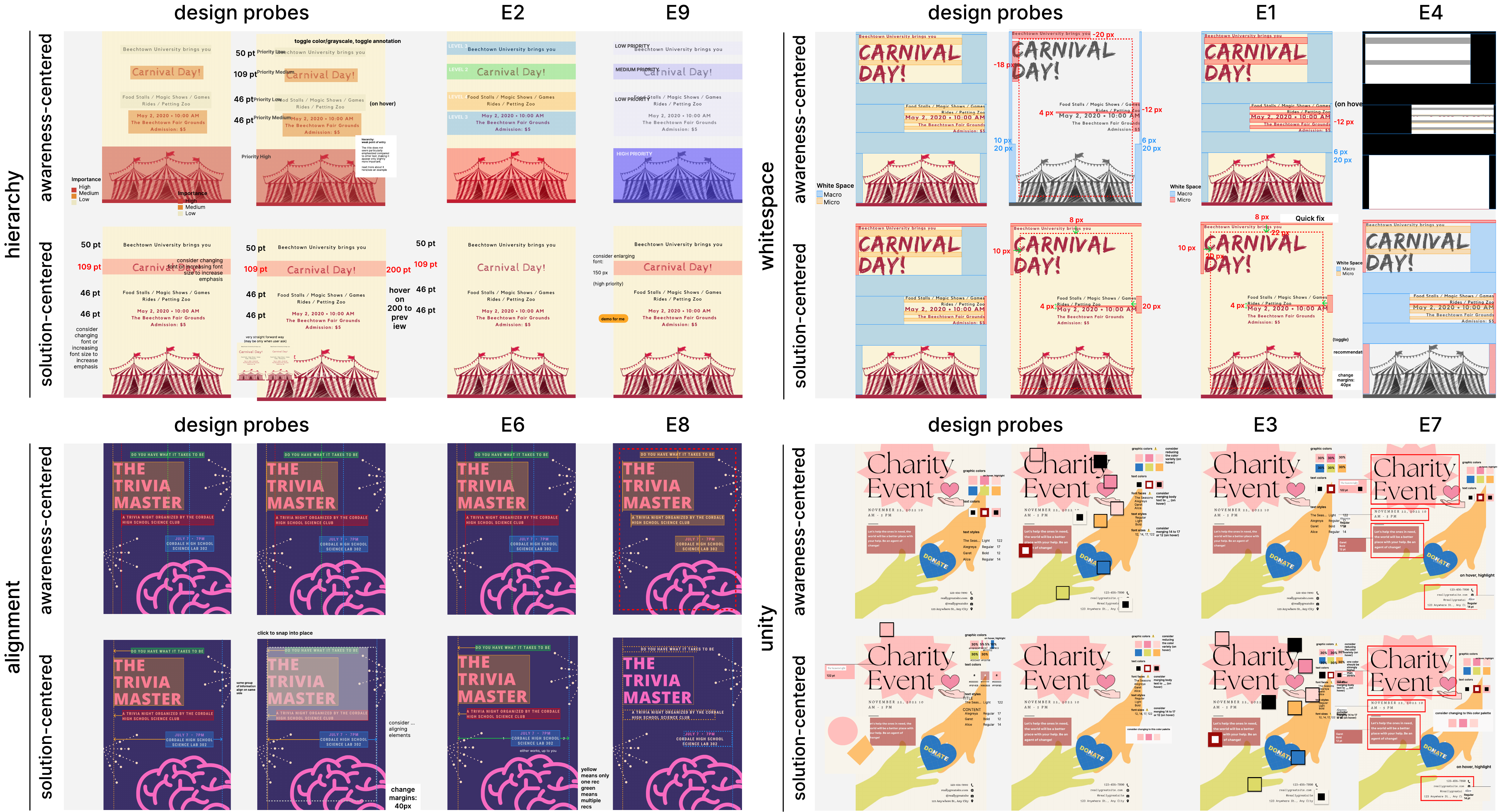}
    \caption{
    Our design probes with a set of four annotations for each design principle (hierarchy, whitespace, alignment, unity). Participants were asked to iterate on these probes based on their own feedback-giving style. The examples shown here are design annotations resulting from their iterations: experts used Figma's drawing and editing tools to annotate, adjust layouts, and propose alternative arrangements while explaining their rationale.}
    \Description{
    This figure shows the design probe. For each principle, it contains 2 awareness-centered designs and 2 solution-centered designs. Next to them, we presented 4 participants' designs. For hierarchy principle, participants use different colors for texts with different weights (awareness-centered) and red text and rectangles to mark title (solution-centered). For alignment, participants use dotted lines, colored boxes to indicate alignment groups (awareness-centered) and arrows and dotted regions (solution-centered). For whitespace, participants used red and grey highlights for texts (awareness-centered) and red margin, green arrows (solution-centered). For unity, participants used red boxes and pixels next to each color.
    }
    \label{fig:designprobes}
\end{figure*}

\subsection{Findings}
We summarize key observations, participant practices, and perspectives on \neutral versus \directed feedback from the co-design study. 

\subsubsection{Experts Fluidly Move Along the Actionability Spectrum} 


Experts emphasized that feedback rarely fell into purely \neutral or \directed categories. Instead, they shifted styles depending on context, using neutral framing to highlight design principles and encourage reflection, and direct suggestions when issues were concrete or time-sensitive. In their freeform feedback, experts also fell along the spectrum (see Figure~\ref{fig:expertfreeform}), with 3 giving solutions (\ref{fig:expertfreeform}.a), 3 identifying issues (\ref{fig:expertfreeform}.b), and 2 providing a mix of identifying issues or surfacing ideas to consider (\ref{fig:expertfreeform}.c).

One participant reflected on the tension between efficiency and agency, noting both that ``\textit{sometimes you do want direct feedback like when you're on a crunch and need a fix right away}'' (E1), and that ``\textit{not giving this right or wrong answer is important, otherwise automated feedback could dictate creativity}'' (E1). Many described combining the two, beginning with neutral prompts and escalating to directed recommendations when problems persisted: ``\textit{first give neutral feedback, they might be able to identify the issue, and if it's still a problem, give directed feedback}'' (E7).

\subsubsection{Annotations as a Core Communication Medium} Visual annotations were a central way for experts to communicate critiques, further validating our visual annotation approach. While some solely used comments, most participants used arrows, circles, or highlights to draw attention to specific areas, often embedding explanations directly on the canvas. These marks were paired with structured mechanisms, such as sketching guide lines (e.g., to show a margin), displaying font sizes, or outlining alignment groups, to help novices better visualize the underlying design parameters and concepts. Annotations were often complemented with brief textual clarifications that allowed participants to connect what they saw from annotations with the reasoning behind them. As one participant explained, combining annotations with text made it easier to ``\textit{judge if I want that to be the case or if it's completely off base}'' (E2).

\subsubsection{Expert Perspectives on \Neutral Versus \Directed Feedback}
Across principles, experts viewed \neutral annotations as lighter in tone and color, mainly serving to raise awareness or prompt reflection. By contrast, \directed annotations were associated with more explicit cues, often using strong colors like red and green to indicate problems and solutions. Several emphasized that \neutral designs provided space for creative interpretation, whereas \directed ones were appreciated for efficiency in straightforward cases. 

During the initial interview (prior to researchers introducing the concepts), E2 mentioned the difference between what he called ``\textit{critique and direction},'' which corresponded to our \neutraldirected concepts. He clarified, \neutral critiques should highlight what is happening, while \directed goes further to “\textit{tell you how to fix it}” (E2). He noted that in practice, “\textit{some (learners) need more critique than others. Some need a lot of direction because they just get stuck… some students don't need that}” (E2), showing how the balance between \neutraldirected feedback shifts with learner experience. While providing critiques, he provided various ideas for the same concept (see bottom-left of Figure~\ref{fig:expertfreeform}.c), illustrating that there can be multiple appropriate annotations for the same principles.


\begin{figure*}[th]
    \centering
    \includegraphics[width=\linewidth]{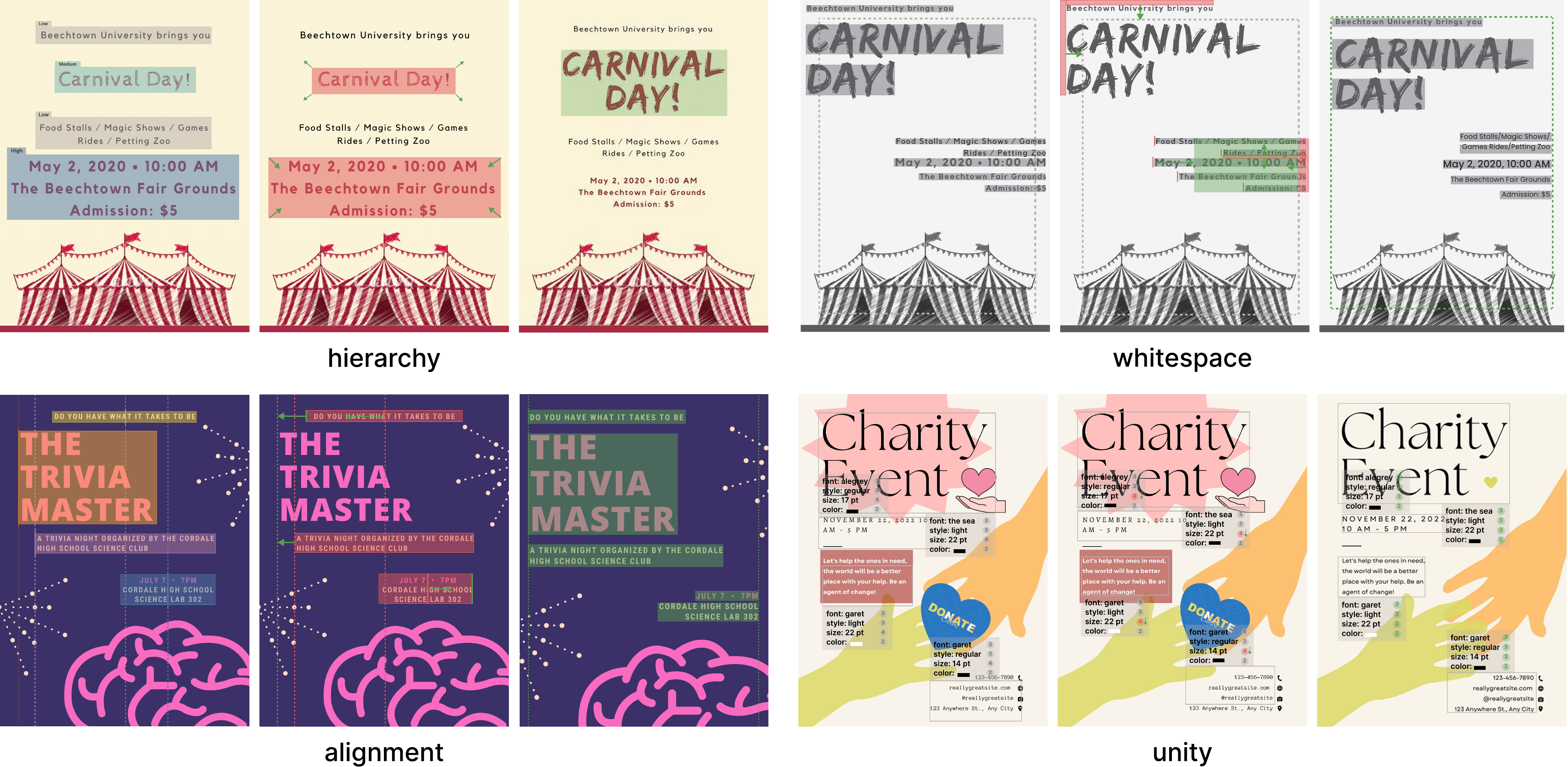}
    \caption{Final set of annotation designs across hierarchy, whitespace, alignment, and unity. For each principle, annotations are shown left to right: \neutral, \directed with issue(s), \directed with no issues.
    }
    \Description{This figure shows our final set of designs. For awareness-centered annotation, we used different colored text boxes for different text elements or text elements in different hierarchy or alignment groups. Grey text and whitebackground for whitespace and a little table next to the text listing properties for unity. For solution-centered annotation, we used red to indicate issues and green (text boxes or arrows) to indicate suggestions. For solution-centered annotations with no issues, we either highlight key elements in green or the minimum margin green (for whitespace).}
    \label{fig:annotation-design}
\end{figure*}

\subsection{Insights on Annotations and Actionability}

Experts confirmed that annotation styles and mechanisms needed to be individually tailored to each design principle. Here we describe some of the lower-level annotation design ideas from the experts, and introduce our final annotation designs by principles. Figure~\ref{fig:designprobes} shows a set of our design probes and expert proposals, and Figure~\ref{fig:annotation-design} shows our final set of annotation designs. 
\revi{Since our goal was to embed annotations computationally into a visual design tool, we applied} a consistent visual language across \revise{different principles}.

After rounds of iteration, we removed readability as a principle, as visualization ideas drastically differed from other principles and there were various existing accessibility checkers for issues like color contrast \cite{contrastcheckWebsite}. We also focused primarily on authoring the text elements as they were 
most directly related to the principles and most emphasized by experts' feedback designs. However, we envision this as an extensible platform where instructors can add custom annotations for additional principles and visual elements. 


Our annotation designs (\revi{see Figure \ref{fig:annotation-design}}) use consistent language in how elements are annotated. Other than necessary exceptions, text elements are highlighted with filled rectangles and dotted lines indicate guide lines in the canvas (such as for alignment or margins). 
For \neutral annotations, text elements are highlighted to visualize concepts relating to each design principle. Red and green are intentionally avoided to minimize potential misinterpretation of something as ``correct'' or ``incorrect'' due to their preexisting associations. 
For \directed annotations, red and green denote problematic regions and possible solutions: red boxes indicate an issue and green arrows suggest a possible action to fix it.
When there is no issue, green is used to highlight elements related to the principle. \revise{All colors come from the Tableau color palette \cite{tableauWebsite}, chosen based on feedback from co-design experts, who emphasized the importance of neutral, non-ambiguous colors.}

\subsubsection{Hierarchy} Experts visualized text parameters to show relative importance, using font-size listings and overlays that previewed how larger titles would shift emphasis. 

\begin{quote}
    \emph{\Neutral}. Annotations use color (shades of gray to blue) to visualize levels of visual emphasis (high, medium, low). This shows how visual attention is distributed across the design. \\
    \emph{\Directed}. Red highlights flag elements with misaligned emphasis, and green arrows suggest size adjustments to better match the level of importance (e.g., enlarging a title or reducing competing text).
    
    \emph{No issues}. When emphasis aligns with content importance, no issues are flagged. Correctly emphasized elements, such as the title, are confirmed in green.
\end{quote}

\subsubsection{Alignment} Experts preferred structural overlays such as grid lines, margin guides, or colored grouping highlights to expose inconsistent axes, which they considered precise and easy for novices to interpret. 

\begin{quote}
    \emph{\Neutral}. Annotations show \revise{external} alignment across multiple elements (grouping by shared edges) and \revise{internal alignment} within single text element (left, right, center, justified).\\ 
    \emph{\Directed}. Red highlights mark alignment inconsistencies, while green arrows suggest consolidating groups or matching internal alignment with external positioning. For example, the annotations could recommend left-aligning multiple text elements \revi{to reduce the number of alignment groups} or \revi{making right-positioned text be right-aligned internally}.\\
    \emph{No issues}. When alignment is consistent, no elements are flagged. Correct groupings and alignments are shown in green.
\end{quote}

\begin{figure*}[t]
    \centering
    \includegraphics[width=\linewidth]{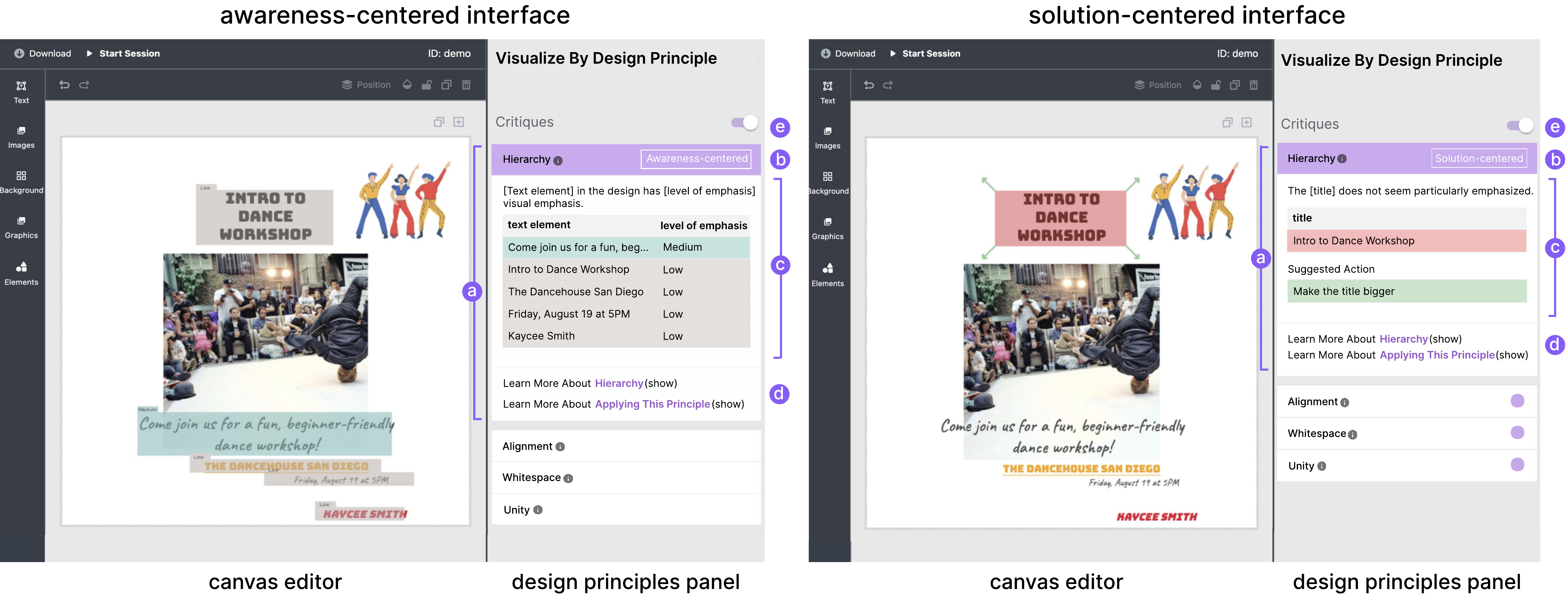}
    \caption{\revise{\textit{VizCrit} interface}: pictured here with hierarchy annotations, 
    \revi{\neutral} on the left and \revi{\directed} on the right. The interface is primarily split into two components: the \emph{Canvas Editor} where users design and see their situated visual annotations, and the \emph{Design Principles Panel} with further text information about the principle and annotation: (a) provides further information about the principle annotation and is expanded by selecting (b), (c) provides a text explanation of the annotation, and (d) provides additional \baseline information about the principle. For \directed annotations, users can use (e) to toggle on and off the purple dot notifications that indicate when there is an issue for a given principle.
    }
    \Description{This figure shows VizCrit interface. On the left, there is a canvas allowing users to do designs by adding or adjusting design elements. There is currently a solution-centered hierarchy feedback, which uses red text box and green arrows to indicate enlarging the title. Next to the canvas is the explanation text. On the right, another awareness-centered annotation shows different colored priority levels of the text along with the explanation text on the right panel.}
    \label{fig:interface}
\end{figure*}

\subsubsection{Whitespace} Experts recommended margin indicators and highlighting the space between elements, as shown in many of their iterations on our designs (Figure~\ref{fig:designprobes}). Some suggested simplifying annotations by removing pixel values, 
while many proposed using grayscale transformation to contrast positive and negative spaces. 

\begin{quote}
    \emph{\Neutral}. Text is abstracted into gray boxes against white background, highlighting the empty space between elements and margins. Dotted lines indicate suggested boundaries for spacing. \\
    \emph{\Directed}. Red highlights mark insufficient or uneven spacing, while green arrows or boxes suggest increasing margins or evening out line breaks. For example, the annotation could recommend shifting elements inward to create more consistent spacing. 
    
    \emph{No issues}. When spacing is sufficient and consistent, no issues are flagged. Margins and gaps are confirmed in green.
\end{quote}

\subsubsection{Unity} Experts emphasized embedding parameters into the design, for example, generating color palettes, showing percentage breakdowns of area covered by different hues, and overlaying variations in fonts and styles to make inconsistencies more visible.

\begin{quote}
    \emph{\Neutral}. Each text element is annotated with its font family, style, size, and color, along with counts of distinct values. This shows how many variations exist in the design. \\
    \emph{\Directed}. Counts are highlighted in red to flag excessive variation, while green arrows suggest reducing counts (e.g., limiting the number of total font sizes) to strengthen consistency. 
    
    \emph{No issues}. When the number of variations is within acceptable limits, no issues are flagged. Current counts are shown in green to confirm visual cohesion.
\end{quote}

\section{\systemname System}\label{sec:system}

We introduce the design and implementation of \revi{\textit{VizCrit}\footnote{VizCrit Code and other study materials can be found on our project page: https://ejane.me/vizcrit.html}}, a visual design tool that automatically generates and displays feedback as annotations (see Figure~\ref{fig:interface}).
We describe how we translate instructor critiques into computational heuristics that can generate design annotations, detect issues, and suggest solutions in real time. \revise{Our current implementation relies on heuristics-based algorithms rather than generative models, because we aim to provide granular and interpretable annotations grounded in principles and annotation practices from design instructors and professionals.}

\begin{figure}[t]
    \centering
    \includegraphics[width=\linewidth]{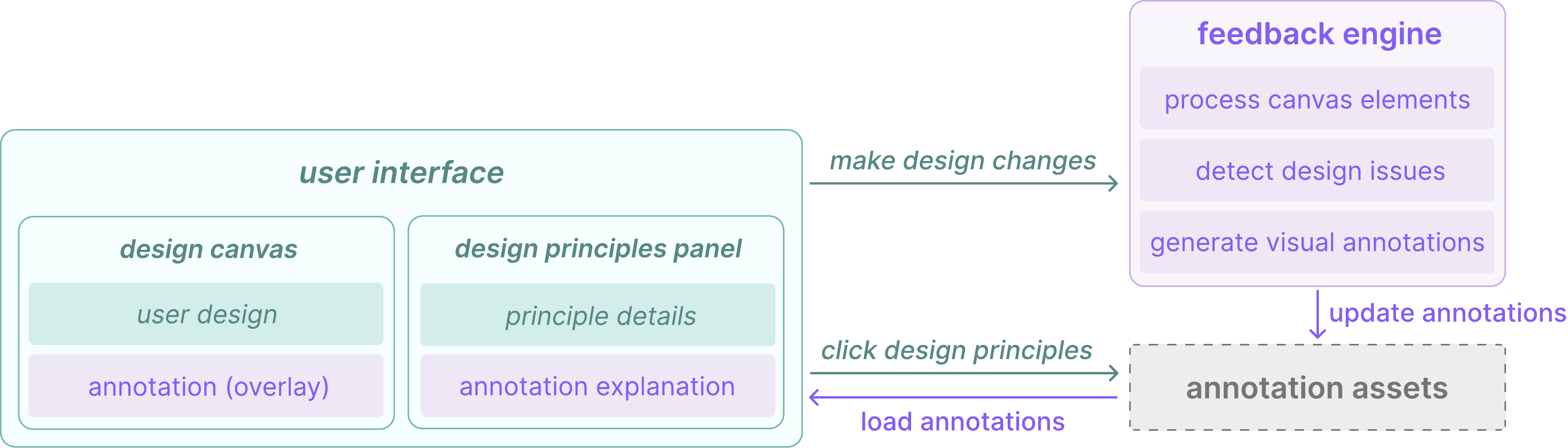}
    \caption{\revise{A system diagram} from the user's perspective. The purple color indicates interactions with feedback and design annotations, and the cyan color \revised{with italic text style} indicates user interactions. Whenever the user makes any design changes on the canvas, the frontend (interface) will send a request along with the current design state to the server (feedback engine). The feedback engine then analyzes the design and makes updates to the annotation files in annotation assets. Annotations will be loaded to the interface whenever the user clicks design principles on the principles panel to request feedback. This ensures the efficiency of our real-time annotation rendering.}
    \Description{This figure shows the system diagram from the user's perspective. The left cyan box indicates user interface that contains design canvas and design principles panel. There is an arrow pointing from interface to feedback engine, which processes canvas elements, detects design issues, and generates visual annotations. There is an arrow pointing from the feedback engine to annotations assets saying update annotations. An arrow from the user interface points to the annotation assets saying click design principles. Accordingly, an arrow points from annotation assets to the user interface saying load annotations.}
    \label{fig:system-diagram}
\end{figure}

The system is composed of two main components: a \emph{server} for performing algorithmic computations for \neutraldirected annotations, and a \emph{user interface} that provides the design canvas and surfaces annotations to the user (see Figure \ref{fig:system-diagram}).
The user interface is built on top of Polotno\footnote{https://polotno.com/}, an open-sourced web-based visual design tool and the server is a Python backend with a WebSocket that facilitates communication between them. \revise{A detailed description of how the backend server communicates with the user interface is included in Appendix \ref{sec:implement}.} 

\subsection{\revise{User Interface: Surfacing Annotations in a Visual Design Tool}}
The user interface has two main sections: \emph{Canvas Editor} and the \emph{Design Principles Panel}. The editor allows users to create visual designs using elements like text, images, and graphics. On the right side of the canvas, the panel offers control for real-time and on-demand design annotations.

\subsubsection{Design Principles Panel}

The main portion of the panel comprises expandable areas per principle: Hierarchy, Alignment, Whitespace, and Unity.
The user can click on a principle name (\ref{fig:interface}.b) to expand that section. Upon click, the action is also reflected in the design editor through showing the corresponding annotation as an overlay to the canvas. 
Within the principle, there are two components:
\emph{Annotation Explanation} (\ref{fig:interface}.c) and  \emph{Principle Details} (\ref{fig:interface}.d).

\subsubsection{Annotation Explanation}
\begin{figure*}[t]
    \centering
    \includegraphics[width=\linewidth]{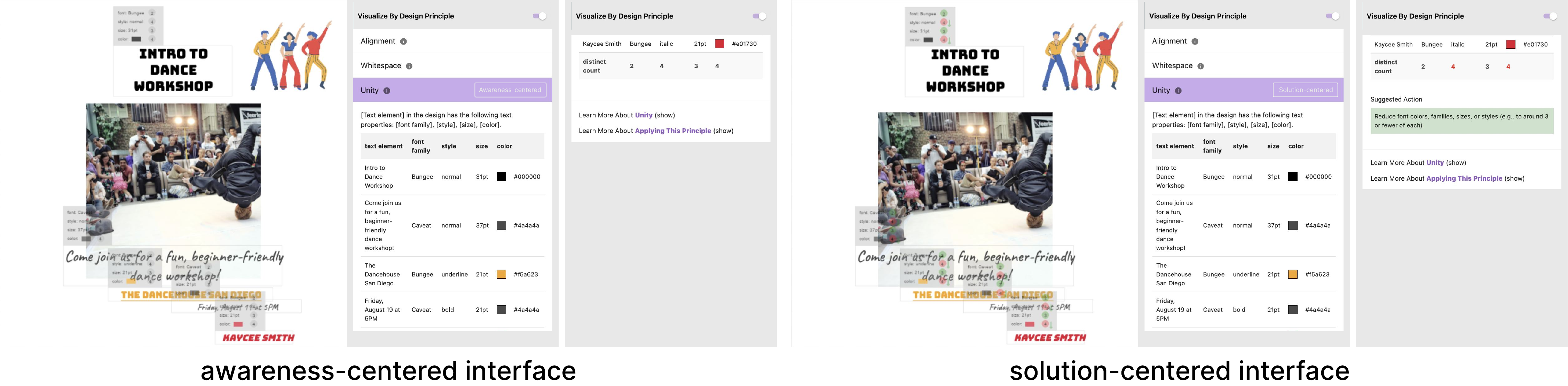}
    \caption{Each of our visual annotation had a corresponding text explanation on the right panel as tables. Pictured here are the unity annotations with their explanations \revise{to help interpret visuals}: \neutral (left) and \directed (right).}
    \Description{This figure shows that for each visual annotation, we include a text-version explanation table. On the canvas, it shows unity annotations, on the right panel, it shows a table listing all the unity text properties including family, style, size, and color.}
    \label{fig:visualandtext}
\end{figure*}

To help users interpret design annotations, annotation explanations (\ref{fig:interface}.c) are provided on the right panel as a text description of the visual annotation in a table format (see Figure~\ref{fig:visualandtext}). It has a consistent color-coding with the design annotation and replicates all visual information into a text representation. These are designed to be as close as possible to a direct mapping so they communicate equivalent information. For the \directed version of annotation, there is an extra ``Suggested Actions'' field that translates the green arrows into their implied actions for fixing design issues. \revise{When the annotation is occluded by the design, the user can also refer to the annotation explanation for more details}. Appendix ~\ref{tab:tabletexts} shows a comprehensive list of annotation explanations.

\subsubsection{Principle Details} Static \baseline references for each design principle (\ref{fig:interface}.d) are provided. To avoid information overload, all additional details are available only after the users click the ``show'' button. The first collapsed information helps in understanding the design principle, including its definition and a common issue. The second collapsed information helps in applying the principle by listing a relevant example and actions that could be taken, such as increasing the prominence of title text to follow Hierarchy principle. This is the design principle-related information available in \textit{\systemname} across all feedback conditions. See Figure~\ref{fig:teaser} and Appendix~\ref{tab:principledetails} for more principle details.


\subsubsection{Critiques Toggle} In the \directed condition, the purple dot next to a principle (\ref{fig:interface}.b) indicates potential design violations of that principle along with feedback. If there are no design issues, the dot will have a lighter color and a check sign. Users can use the critiques toggle (\ref{fig:interface}.e) to turn on/off the notifications. Since issues are only surfaced in \directed annotations, this is hidden in \neutral and \baseline conditions.

\subsection{Server: Generating Annotations Computationally}
We introduce heuristic algorithms for generating annotations for each design principle. For \neutral annotations, this mainly involves analyzing the design elements to generate visual annotations. For \directed annotations, this additionally involves detecting issues and determining reasonable solutions. We describe initial steps for processing text that are applied across multiple principles and then describe the algorithms per principle.

\subsubsection{Processing Text Elements} 

OCR analysis is performed on text elements  to measure their spacing, extracting text structure with OpenCV contour detection\footnote{https://docs.opencv.org/3.4/d4/d73/tutorial\_py\_contours\_begin.html} to obtain their precise pixel-level bounding boxes. This is particularly needed for Hierarchy and Whitespace principles as they require inspection for the space physically occupied by the text.
Then, for each text \revise{element}, 
text line heights are \revise{determined} using HarfBuzz text shaping library\footnote{https://harfbuzz.github.io/}.

\subsubsection{Hierarchy: Examine Visual Prominence} 
To analyze the level of emphasis of text elements, K-means clustering algorithm is performed on the elements' line heights to form two or three different groups. Then, the cluster centers are used to compare with font size thresholds for determining the low, medium, high priority groups. These are directly visualized for \neutral annotations. For \directed annotations, the system additionally  detects issues through identifying a weak (or no) point of entry. If there are no elements in the high priority group, the system will report a ``No elements seem to be visually emphasized'' issue. If multiple elements exist in the high priority group, the system will report non-title texts as competing elements. In both cases, the system will suggest that the user make the title larger. In the second, it will additionally propose reducing the size of the competing element.

\subsubsection{Alignment: Obtain Horizontal Groups}
From the canvas JSON object, text elements are filtered, and their x-coordinates and alignment attributes are used to calculate how elements are horizontally positioned. Next, they are grouped based on their horizontal alignment, and the system will indicate ``too many alignment groups'' issue if several text elements are aligned to different axes. Additionally, the group's width is computed to determine left/right boundaries and to check if their external positions on the canvas match their internal alignment. Issues will be reported if elements are misaligned (e.g., text on the left side of the canvas is center-aligned). In the first case, the system will suggest aligning elements to fewer axes. In the second, it will propose matching the external and internal alignment.

\subsubsection{Whitespace: Perform Spacing Analysis}
The awareness centered annotation directly surfaces the text processing described earlier to showcase the boundaries around text elements and highlight the whitespace between them.
For \directed issue detection, three distinct whitespace analyses are performed: canvas margin, inter-element spacing, and ragged text edges. To detect whether elements are positioned too close to the canvas edges, we set a minimum threshold based on the canvas size. If below the threshold, the system will direct the users to move elements farther from the edge.
\revise{The detection of insufficient spacing between elements involves a pairwise analysis. First, filtering logic is conducted to exclude elements that are far apart.} Next, elements are compared if they are horizontally and vertically adjacent to one another. 
The system will suggest moving two text elements farther apart from each other.
Ragged edges occur when a text block has notably different lengths in each line, creating visual disruption for the reading flow. After obtaining the widths from the OCR analysis, 
the shorter lines are compared with the longer lines to determine whether they are significantly shorter. The system's suggestion will be to even out the vertical lengths.

\subsubsection{Unity: Count Text Variances} 
The JSON objects of text elements are used directly to count the variations in font family, style, color, and size. 
These counts are surfaced per text element in the \neutral annotation. In the \directed annotation, if any text elements have more than three variances for any text properties, the system will show the count in red and indicate a ``too many variances in text'' issue. \revise{In the visual annotation, a down arrow is used as a signal to reduce the number of text properties to three or fewer.}

\section{Technical Evaluation}\label{sec:technicaleval}
Annotation generation, issue detection, and solution suggestion algorithms are evaluated systematically. Among them, issue detection algorithms (for hierarchy, alignment, whitespace, \revise{and} unity principles) are the primary focus.

The evaluation dataset came from a public dataset~\cite{e2024timing},
containing 20 
Polotno Studio participant designs. 
Authors generated 26 additional test cases during the implementation process to capture every issue instance within each principle and particularly focus on the edge cases. 
Algorithms were compared against design expert ratings from the original dataset as ground truth, and \revise{the authors reviewed all test cases for accuracy.} Ambiguous cases (6 hierarchy, 3 alignment, 4 whitespace, 5 unity), \revise{where designs were labeled as issues in the dataset but not by our focused heuristics, or vice versa,} were marked N/A and produced a modified ground truth set. 

Our issue detection algorithm demonstrates effectiveness in identifying design issues, 
achieving > 0.85 (out of 1) accuracy across principles using modified ground truth. 
Here is a detailed breakdown for each principle (expert/modified): Hierarchy: 0.95/1; Alignment: 0.79/0.88; Whitespace: 0.84/0.93; Unity: 0.7/0.93. 
For the annotation generation and solution suggestion, 
three authors visually inspected system-generated annotations overlaid on top of original designs in Miro, producing 152 verified cases. We iterated and refined the heuristics until annotations appeared appropriate across all test designs.





\section{Study on Feedback Actionability with Design Novices}\label{sec:usereval}

Since \textit{\systemname} enables different levels of feedback actionability, it provides an opportunity to study how these differences impact novices' creative processes, particularly the tradeoffs between immediate performance and creativity.
\revise{
We conducted a study to explore RQ2, which we expand here into sub-questions:{
\begin{itemize}
  \item \textbf{[RQ2a]} How do novices engage with \textit{VizCrit}'s feedback and annotations throughout their design processes?
  \item \textbf{[RQ2b]} How do different levels of feedback actionability influence \revi{perceptions of creativity, learning of underlying design principles, and design outcomes?}
\end{itemize}
}}


% ...

\subsection{Participants}
We recruited 36 participants (28F, 7M, 1 prefer not to say; age: 18--35, $\mu = 21.4$) through \revi{UC San Diego}'s SONA system\revise{\footnote{https://www.sona-systems.com/}}, \revise{a participant recruitment and study management tool}. All but one self-identified 
as having little to no experience in visual design (one identified as intermediate).


\subsection{Study Procedure}
\revise{We adopted a between-subjects design where} participants had the same access to the feedback panel within \textit{\systemname}, which offered feedback with varying actionability: \neutral, \directed, and \baseline. 
Participants were provided with an initial seed design to revise that consisted of one image, one graphic element, and five text elements and contained one design issue per principle (i.e., four total violations). 
We counterbalanced the three feedback types and two seed designs\revise{---each participant worked with one type of feedback actionability \revise{and one seed design}.}
The study sessions were conducted over Zoom and we obtained verbal consent from participants to capture a screen recording. The sessions lasted 60-90 minutes and participants received credits through SONA.
\subsubsection{Pre-Test} After completing the demographics survey, participants were given a knowledge test to examine their existing knowledge of design principles. We designed the test following Bloom's Taxonomy~\cite{bloom1956taxonomy}: asking participants to remember, understand, and apply visual design principles.
The test comprised questions with two levels of difficulty: easy (Vocabulary Test) and hard (Principles Application Test). Participants were first given a design principles reference sheet that contained each design principle's name, definition, common issue, and example. 
For the vocabulary test, participants were asked to name the principle that a poster pair addressed, and for the application test, they were asked to identify different design issues at specific locations within a design.
\subsubsection{Tool Tutorial \& Practice Task} Participants were guided to perform a series of editor operations (adding text, changing backgrounds, etc.)
to get familiar with the tool's editing functionality. 
Next, they were provided with an intentionally simplistic starter design with clear violations across four principles as a practice task to address issues \textit{without} feedback.
\subsubsection{Annotation Tutorial \& Design Task} Participants were provided with a document that explained every design annotation that they might encounter in the later design task. 
They could also explore the practice task with design annotations to further familiarize themselves with \textit{\systemname}. Participants were then introduced to the main design task: revise another (more visually complex) seed design to make it visually appealing and fit their creative preferences. \revise{We intentionally provided a shared flawed seed design to control for differences in participants' initial design skills---this helped ensure that all participants encountered at least one issue per principle, making evaluations on participants' design outcomes and creativity more comparable across conditions}. Participants were asked to think aloud and use the task as a learning opportunity to develop their design skills. 
\subsubsection{Post-Test} Participants repeated the knowledge test as a post-task evaluation. The difference between the pre- and post-test captures the degree of learning as a result of the activity.

\subsubsection{Survey \& Interview} Participants completed a survey consisting of the Creativity Support Index~\cite{cherry14CSI} with 5-point Likert-scale ratings that assessed their perceptions of their designs, design knowledge, experience with \textit{\systemname} and annotations, and some AI-usability metrics~\cite{Oh18ILead}. The study concluded with an interview focusing on the participants' overall design and learning experience.

\subsection{Data Collection and Analysis}
\begin{table*}[t]
  \centering
  \caption{\revise{An overview of quantitative results from participants' self-assessed Creativity Support Index (CSI), expert evaluations, feedback interaction analysis, and the learning test. The asterisks denote the level of statistical significance: * indicates $p < .05$, and ** indicates $p < .01$.} \revised{Conditions denote the pairwise comparisons that were statistically significant, and for the comparisons that were not shown in table, they were all insignificant.}}
  \label{tab:quant-summary}

  \small
  \renewcommand{\arraystretch}{1.15}

  \begin{tabular}{p{4.5cm}p{5.5cm}p{2.5cm}p{2.4cm}}
    \toprule
    \textbf{Category} & \textbf{Evaluation} & \textbf{Significance} & \textbf{Comparison} \\
    \midrule

    \multirow{1}{*}{\textbf{Design Quality}}
      & Number of design issues
      & $p < .05$ (*) 
      & Solution-centered $>$\ Textbook-based \\
    \midrule

    \multirow{4}{*}{\textbf{Creativity}}
      & Score from Creativity Support Index (CSI)
      & $p < .01$ (**)
      & Solution-centered $>$\ Textbook-based \\
      & Enjoyment (CSI)
      & $p < .01$ (**)
      & Solution-centered $>$\ Textbook-based \\
      & Results Worth Effort (CSI)
      & $p < .01$ (**)
      & Solution-centered $>$\ Textbook-based \\
      & Creativity score from Expert Ratings
      & non-significant
      & non-significant \\
    \midrule

    \textbf{Degree of Change}
      & Number of design iterations
      & non-significant
      & non-significant \\
    \midrule

    \textbf{Interaction with Feedback}
      & Number of design principle clicks
      & $p < .01$ (**)
      & Awareness-centered \& Solution-centered $>$\ Textbook-based \\
    \midrule

    \textbf{Design Principle Application}
      & Improvement in the learning test
      & non-significant
      & non-significant \\
    \bottomrule
  \end{tabular}
\end{table*}
We describe the data collected during the studies across knowledge tests, design tasks, surveys, and interviews and how we analyze the data.

\subsubsection{Qualitative Analysis} 
To understand participants' perceptions of \textit{\systemname}'s feedback, we conducted qualitative analysis on the interview data.
Three authors open-coded all the interviews on Condens\footnote{https://condens.io/}, following a reflexive thematic analysis method \cite{braun2021thematic} and taking a combination of inductive and deductive coding approaches. 
Codes captured participants' reactions and perceptions of annotations/principle panel 
and their desired improvements for the tool.
Authors met regularly to discuss data and codes, which were then aggregated on a whiteboard to generate final themes. 

\subsubsection{Behavioral Data}
\label{behavorial-data-analysis}
We analyzed participants' interaction with \textit{\systemname}, including their responses to annotations and design principle clicks, along with the evaluation on \textit{\systemname}'s feedback quality.

\paragraph{Response to Annotations} 
To further capture how participants responded to design annotations, whether they immediately accepted the system's suggested solutions (in \directed condition) or made relevant changes to address design principles (in \neutral condition), and which design iterations 
were directly influenced by \textit{\systemname}'s feedback, we coded participants' subsequent design changes after they viewed annotations.

\paragraph{System Usage Logs and Analysis} 
We logged participants' interaction data using Firebase Firestore Database\footnote{https://firebase.google.com/}. The logs captured 1) principle clicks (timestamp and principle name), 2) every design annotation viewed (either when a principle panel was opened or when annotations were refreshed within an open panel), and 3) participants' on-canvas design states. Both design annotations and participants' designs were stored as JSON. The principle clicks data was cleaned \revi{by removing any clicks outside of the design task} and visualized for analyzing the total number, frequency, and timing of feedback requests.

\paragraph{Feedback Quality}
\revise{Our technical evaluation focused on accuracy for a test set. To further evaluate the feedback quality actually seen by participants, we randomly selected three design annotations that each participant viewed among their early, middle, and late feedback requests. Two authors with design expertise rated on the accuracy of annotations (0: inaccurate, 0.5: partial, 1: accurate). }

\subsubsection{Quantitative Analysis} 
Since our study involved three different types of feedback, 
we performed statistical analysis using Kruskal–Wallis H test~\cite{kruskal1952use}. If the Kruskal-Wallis test reported a significant effect, we then conducted pairwise post-hoc comparisons using Mann–Whitney U test~\cite{mann1947test} with Bonferroni correction. For comparisons 
\revise{only involving} two independent groups (e.g., \neutral versus \directed annotations), we also used a Mann–Whitney U test~\cite{mann1947test}. Given the sample size, \revise{some comparisons were not statistically significant \revi{(see Table \ref{tab:quant-summary})}. We mark these all as `non-significant' in our reporting.}

\paragraph{Self Assessments}
We administered the Creativity Support Index \cite{cherry14CSI} but marked the `Collaboration' dimension as `Not Applicable' (i.e., 0 in the 0-10 scale) for all participants for consistency, since participants completed the design task individually. 

\begin{figure*}[t]
    \centering
    \includegraphics[width=\linewidth]{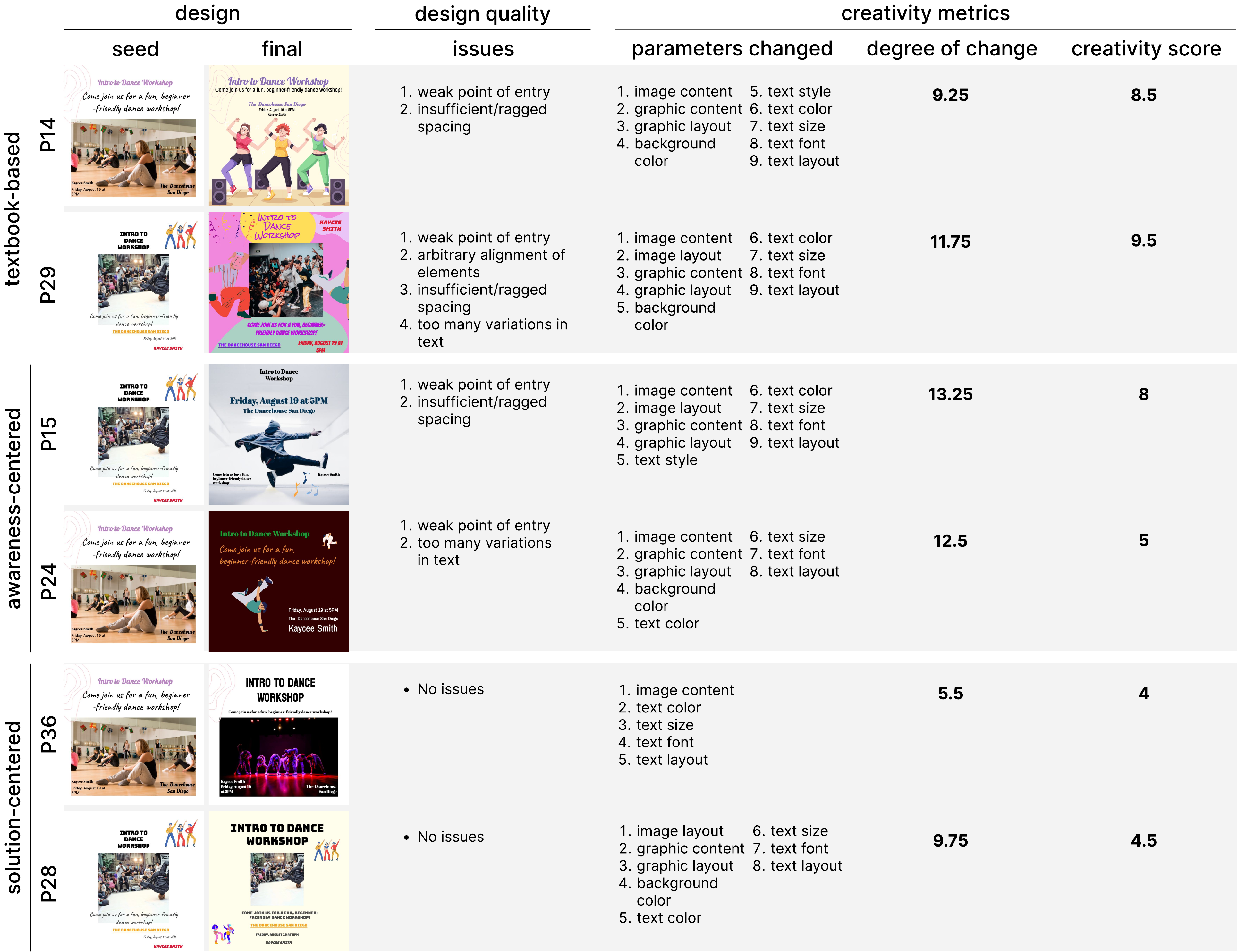}
    \caption{Example participant designs and expert ratings across conditions. Experts annotated design issues and parameter changes, and provided ratings of creativity (1: low, 10: high) and degree of change relative to the seed design.
    \Description{This figure shows a grid showcasing participant designs across conditions and some expert design quality and creativity metrics, top to bottom, 2 per condition: textbook-based, awareness-centered, and solution-centered. For each row of designs, it shows first the initial seed design and then the participant's final design.}
    }
    \label{fig:participantdesigns}
\end{figure*}

\paragraph{Expert Assessments}
\revise{To evaluate the participants' design outcomes, we also wanted to bring in a more objective perspective from experienced designers. Since several of the co-authors have significant expertise in design, we performed the following \revised{author-based expert evaluation} (see Figure \ref{fig:participantdesigns} for examples). All evaluations were individually graded, reconciled, and then averaged.}

\begin{itemize} 
\item{\textbf{Degree of Change.}}
We measured the degree of change between participants' final designs and initial seed designs to understand the extent that they iterated on their design. Two authors developed a design iteration metric based on \textit{each} canvas element: images (content, layout), graphic elements (content, layout), text elements (style, color, size, font, layout), and background color.
For the images and graphic elements, authors coded participants' design iterations in a randomized order
and reviewed the results together to resolve any inconsistencies.
The text elements and background color iteration were programmatically compared through design JSON files.
\item{\textbf{Creativity.}}
We used a subjective score for the creativity of final designs (10-point, 1: bad, 10: good).
\item{\textbf{Design Quality.}}
We measured design quality by counting the number of design principle violations. Two authors with design expertise rated the number of issues in participants' final designs. We developed 4-point scoring rubrics covering four design principles.
\end{itemize}

\subsubsection{Knowledge Test (Learning)}
Two authors with design expertise developed 16-point scoring rubrics and graded pre- and post-knowledge tests. After grading individually, we reviewed any outstanding scores that had differences greater than one. We revisited these results and reconciled the differences. All scores were then averaged. Then, we measured learning 
by computing the differences between the post-test and pre-test scores. 
We analyzed the results separately for Vocabulary (easy) and Principles Application (hard) sections to capture more nuanced learning outcomes. 

\subsection{Results}\label{sec:results}
In this section, we 
describe how participants engage with feedback, their perceptions of visual annotations, and the impact of different levels of feedback actionability on \revi{ process-related behaviors, learning of design principles, perceptions of creativity, and overall design outcomes. Results are organized by sub-questions of \textbf{[RQ2]}}.

\subsubsection{
RQ2a: How do novices engage with VizCrit's feedback and annotations throughout their design processes?
}

\paragraph{Motivations and Timing for Requesting Feedback}
Participants across different feedback actionability conditions demonstrated distinct motivations and behavioral patterns for accessing \textit{\systemname}'s feedback.
To analyze at what times and design phases participants sought feedback, we divided them into ``in-action'' participants who periodically requested feedback and ``on-action'' participants who mainly engaged with feedback at a later stage after completing a substantial amount of work \cite{e2024timing}. For example, P30 (solution-centered) \revi{who was ``in-action''} reflected that she checked annotations throughout the design process: 
``\textit{anytime after I adjusted something. I just wanted to make sure everything was lined up.}'' For P20 (solution-centered), in contrast, exemplified ``on-action'' participant who only used annotations to ``\textit{fine tune}'' his design. 

Most \baseline participants \revise{were on-action} who checked feedback only at the end (on-action versus in-action in \baseline = 8:3, N/A: 1; \neutral= 4:8; \directed= 6:6). Indeed, both \neutraldirected participants requested significantly more feedback than \revised{\baseline participants} (\neutral: $ M= 12.83, SD= 7.65$; \directed: $M=9, SD= 4.59$; \baseline: $ M= 3.91, SD= 1.76$; $p<.01$).  Since \baseline participants only received generic and static educational content and they had already gained some design principle knowledge from the pre-test reading, they found less value in referencing principles repeatedly. As P12 stated, ``\textit{I don't think I was using it so much throughout the process... at the end I looked through it to review that I had done everything. I hadn't forgot anything.}''
Therefore, for these on-action participants, design annotations or principles became like a final checklist to help identify remaining issues and validate participants' design choices.

Participants treated \textit{\systemname}'s feedback not only as a checker, but also as a visual inspection tool that offered insights into unnoticed details in \neutraldirected conditions.
Design annotations augmented participants' visual ability to locate issues that would otherwise be impossible to find. As P7 (\neutral) commented, ``\textit{I was able to see exactly what parts were not looking the best and where I should put them... the whitespace one was the most convenient because I could actually see the borders and the margins and stuff cause my eyeballing is very bad}''
Many participants also used design annotations to find inspiration or to get unstuck. For example, P20 (\directed) mentioned that he previously wanted to center-align all text elements, but then \textit{\systemname} offered a left alignment suggestion, and he decided to adopt a new layout that aligned elements only on the left and right sides. He said, ``\textit{it (\systemname) helped me explore different things that I wouldn't have thought of previously. And I really appreciate that.}'' Relatedly, P15 (\neutral) reflected using \textit{\systemname} when he was hesitant about where to place new elements: ``\textit{if I'm struggling with seeing where to add or where to place a text, I would use a design tool as (it) can help me with more direction.}''

\paragraph{Some Participants Desired More Actionability}
Several participants expressed a desire to \revi{move even further along the actionability spectrum and shift away from reflective thinking toward productivity-focused moves.}
They wanted an automated system that could help them resolve issues more easily and efficiently. For example, P1 (\neutral) wished to have a click-and-apply feature that reduced the effort of self-identifying and addressing issues: ``\textit{you could just click on each principle and it was just automatically applied to it. And then you can see which one looks better or something. And then you could just like keep it.}'' Similarly, P24 (\neutral) requested performing a simple action on the right panel, not the canvas, to make texts more consistent. She suggested, ``\textit{there's a different type of font and everything, click and then make all this (different) font to one same font from here (the right panel)}.'' Rather than engaging with their own designs, participants were ready to hand some creative agency to the system to achieve better and quicker design results.
In addition to automatically fixing issues, some \neutral participants also requested more direct hints and guidance (similar to our \directed feedback) to reduce their effort in self-evaluation. As P22 proposed, ``\textit{a check to see if everything looks good... 
So hierarchy is done and hierarchy is checked or something like that to see oh that's good}.'' \revise{This aligned with the predictions from our co-design experts who thought \directed annotations could accelerate the design process, but there should be a balance between efficiency and creative ownership.}  Participants' requests reflect an over-simplification of the iterative design process and entirely turn it into a series of automated check-click-apply steps that hinder creative exploration, agency, and ownership.

Participants further requested mechanisms to measure creativity or visually appealing aspects of a design. They requested ``\textit{a visual appeal score}'' (P1) and  ``\textit{some annotations on the creativity part or on the trends part}'' (P2).
This may indicate that novices could have more trust in the computational tool to the point that it overrides their own intuition, judgment, or even creative thinking abilities. 
\revi{Sliding too far towards full actionability and entirely trusting the tool may lead novices to over-reliance and limit learning opportunities, as they experiment less and reflect less critically on their own choices.}

\begin{figure*} 
    \centering  \includegraphics[width=.8\linewidth]{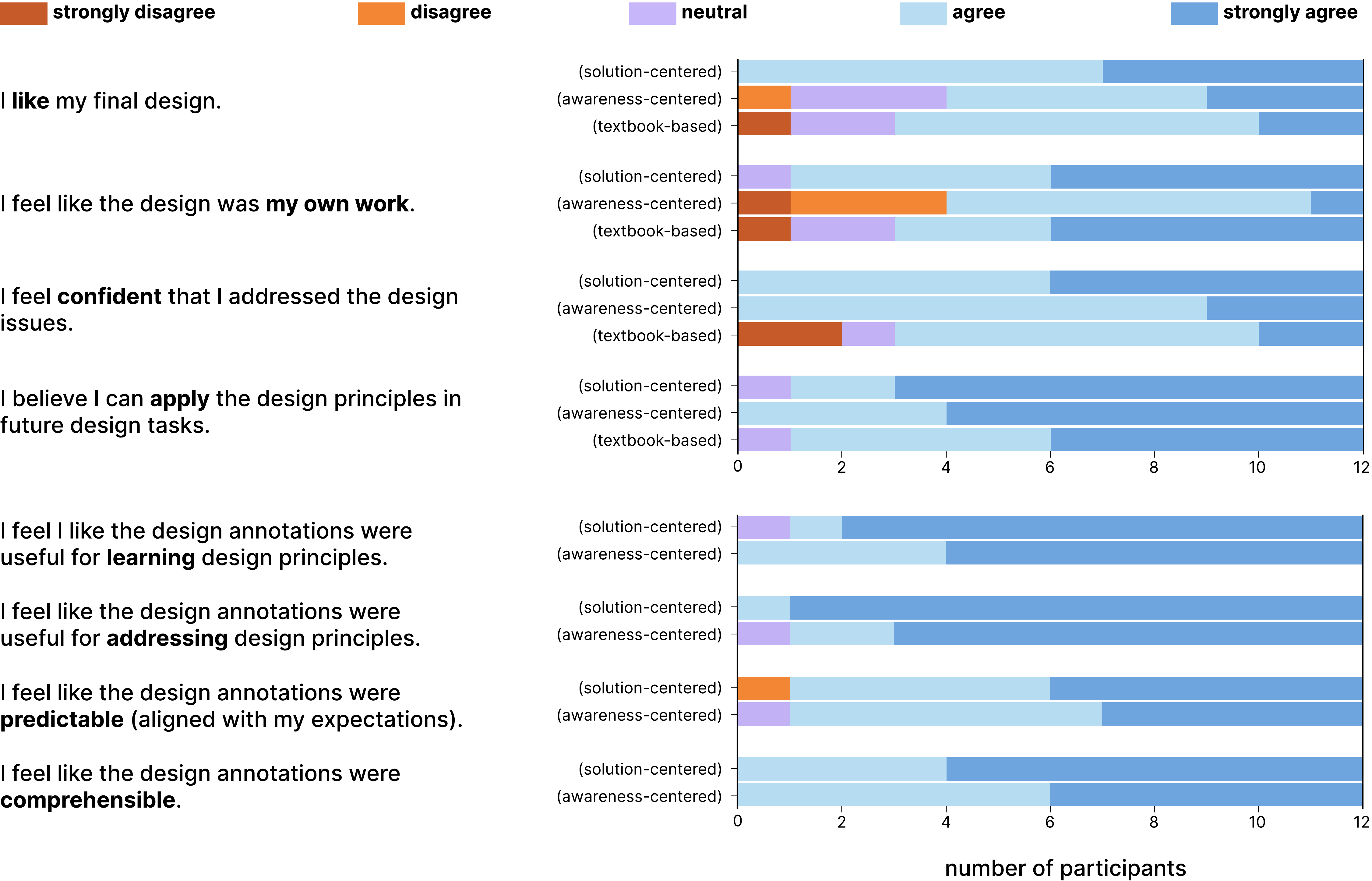}
    \caption{Distribution of participants' ($n=36$, 12 per feedback condition) self-ratings on their final design, creative ownership, confidence, design knowledge, and overall experience with design annotations (awareness-centered or solution-centered). There were no statistically significant differences in these ratings. \revised{However, the ratings on whether the design felt like ``my own work'' seem to contradict our hypotheses about creative ownership. While we expected \directed participants to have the least ownership, they felt it equally as much if not more compared with other feedback conditions.}}
    \Description{
    This figure shows bar charts of likert scale survey responses for: I like my final design, I feel like the design was my own work, I feel confident that I addressed the design issues, I believe I can apply the design principles in future design tasks, I feel like the design annotations were useful for learning design principles, I feel like the design annotations were useful for addressing design principles, I feel like the design annotations were predictable (aligned with my expectations), I feel like the design annotations were comprehensible. Across all conditions and participants, the only disgrees/strongly disagrees were for awareness-centered (1) and textbook-based (1): I like my final design, awareness-centered (4) and textbook-based (1): I feel like the design was my own work, textbook-based (2) I feel confident that I addressed the design issues, and solution-centered (1): I feel like the design annotations were predictable (aligned with my expectations). All other ratings were neutral/agree/strongly agree.
    }
    \label{fig:bar-graph}
\end{figure*}
\paragraph{Awareness- and \Directed Annotations Were Effective and Comprehensible.}
In both \neutraldirected conditions, participants found that visual annotations were easy to use and understand, simple, straightforward, responsive, and effectively communicated key principle information. Particularly, all \neutraldirected participants (N=24) agreed on the comprehensibility of \textit{\systemname} annotations (see Figure \ref{fig:bar-graph}), appreciating that the tool was ``\textit{super cool}'' (P3), ``\textit{very easy and just made sense}'' (P20), and \revise{``\textit{very accurate}'' (P24)}.
\revise{Our expert ratings on the feedback quality also reflected high accuracy \revised{for both \neutraldirected annotations}
(solution-centered: 89\%, awareness-centered: 92\%, among selected annotations)}.

One primary reason that participants found annotations comprehensible was that \textit{\systemname} applied effective color-coding to not only text elements, but also their corresponding explanation text on the right panel. This way, \directed participants could easily locate issues and solutions by examining red and green highlights and \neutral participants could easily differentiate elements' level of emphasis or alignment groups by checking the group color. As P17 noted, ``\textit{I like the color coding again and the dotted lines. I think they're helpful visually because I'm a visual learner.}''

Furthermore, participants enjoyed design annotations that were situated and specific to their own designs. Since the annotation also quickly refreshed based on their design changes, participants found such frequent updates effective. P35 (\neutral) mentioned that ``\textit{it's not just telling me what it (design principle) is, it's actually giving me the info relative to my design currently.}'' P16 (\directed) stated that ``\textit{when I changed something, it also changed really fast, which was good to keep up}'' and P9 (\directed) further added ``\textit{getting live feedback on what your actions were doing ... I can like test out my ideas and see how the computer would react to it, which was really nice}.'' This indicated high interactivity and usability of \textit{\systemname}, since participants not only passively received critiques, but also proactively engaged with annotations to explore different creative possibilities.
For \baseline participants who did not receive visual annotations, they requested more visual aids, examples, or on-canvas highlights that could help them better understand or apply design principles (P13, P23, P25, P29, P31). For example, P23 proposed having a more situated visual indicator that helps users better align their canvas elements: ``\textit{show some lines to see what's being aligned or out of place... editor could show different shapes of sections of the page to (help users) see if everything is aligned well.}'' This resonated with \textit{\systemname}'s visual annotations that offered in-context visual signals for evaluating and improving the design. 

However, participants still identified areas for improvement. They found some design annotations hard to interpret, especially for the \directed internal alignment annotation (see Figure \ref{fig:annotation-design}), where \textit{\systemname} used arrows to suggest changing the internal text alignment. Participants misunderstood it as horizontally moving the element. \revise{To address this, \textit{\systemname} should provide a clearer annotation explanation to clarify the concept of internal/external alignment. Also, a different type of arrow could be used to represent different meanings (e.g., concrete arrows for moving elements, dashed arrows for changing alignment)}. 
\revised{In our expert evaluation on feedback quality, although there are no inaccurate annotations, some received partial scores (0.5/1) due to the pre-defined thresholds in our heuristics-based algorithms (e.g., size or distance thresholds for differentiating priority groups in hierarchy or alignment groups).}

Moreover, while mostly efficient, updates to design annotations could experience inconsistent delays due to the cloud rendering service that converted canvas JSON to images.
As a workaround, some participants chose to wait or work on other parts of their designs during the refresh, but for participants who wanted immediate feedback after an iteration, this caused confusion and frustration.

\paragraph{Annotations Enhanced Confidence}
\revi{Many} awareness- and solution centered participants commented that they \revi{gained confidence} after receiving feedback annotations.
Participants considered \textit{\systemname} as ``\textit{an objective standard}'' (P3, \directed) that could reflect how audiences perceive their design.
Thus, when \textit{\systemname} reported no issues or aligned with participants' self-evaluation, participants gained confidence as their designs were objectively validated. 
P33 (\neutral) explained, ``\textit{with eyeballing you can kind of make mistakes... but with this tool, it helps to make sure confidently everything's aligned, everything is visually going to be appealing... it just helps boost confidence (so) you don't have to double check.''} The computational inspection and external validation from \textit{\systemname} helped participants build confidence in their design outcomes.

Furthermore, having an objective inspection could foster collaboration and bridge peer communication. P3 recalled an example of her roommate, who thought designs from collaborators were ``\textit{just not right, design-wise, conceptually,}'' but she did not know how to communicate her frustration and concerns.
Therefore, P3 felt ``\textit{something like this is really helpful and diffusing that tension and making design principles more accessible and known.}'' In this case, \textit{\systemname} could be a mediator of design conversations, which quantified and visualized design issues using established standards. It provides references for peer discussion and smooths the tension of providing and receiving design critiques.

\subsubsection{RQ2b: How do different levels of feedback actionability influence perceptions of creativity, learning of underlying design principles, and design outcomes?}
\paragraph{\Directed Feedback Inflated Self-perceived Creativity}
Participants with \directed feedback felt significantly more creative than \baseline participants (\directed: $M= 83, SD= 12.06$; \neutral: $ M= 71.61, SD= 18.78$; textbook based: $ M= 65.08, SD=12.86 $; $p<.01$). \revised{They enjoyed using \textit{\systemname} more} ($p<.01$) and felt that their designs were more worth the effort than \baseline participants ($p<.01$).
\revised{
However, despite participants' stronger sense of creativity, 
the expert ratings showed no significant differences on creativity} (\directed: $M= 5.04, SD= 2.47$; \neutral: $ M= 4.96, SD= 2.02$; \baseline: $ M= 7.04, SD= 2.12$, \revise{non-significant}) \revised{or the degree of change across conditions} (\directed: $M= 3.17, SD= 1.39$; \neutral: $ M= 3.44, SD= 1.35$; \baseline: $ M= 4.12, SD= 1.18$, \revise{non-significant}).

\revised{To contextualize this misalignment, we conducted an exploratory analysis of participants' behavioral responses to annotations.} 
Of all design annotations viewed, \revised{the immediate feedback adoption rate \footnote{We measured the immediate feedback adoption rate as the proportion of viewed feedback instances that were followed by a design change that acted on that feedback (see Section \ref{behavorial-data-analysis}, Response to Annotations).}} was 82\% for \directed \revised{participants and} 42\% for \neutral participants. 
We also observed several concrete instances \revised{suggesting that \directed participants felt an urge to implement \textit{\systemname}'s feedback}
even when they felt confused. 
Taking the `weak point of entry' issue as an example, 
after viewing annotations, P6 (\directed) increased the title size but the alert persisted. Although surprised, she continued following the feedback instead of  questioning \textit{\systemname} or considering alternative solutions:``\textit{bigger? OK, maybe I'll make this (size) 55.}'' She felt compelled to act and even prepared to make title larger again while the new annotations were loading. After seeing the updated annotation indicated no issue, she immediately moved her cursor away from the title.
For \neutral participants, however, they better retained their creative agency and were more likely to stop iterating even if the annotations did not completely align with their expectations. For instance, P33 
began by making the title bigger and then felt unsure about the feedback: ``\textit{I'm making it (title) bigger. It's still saying medium level of emphasis, but I'm making it bigger.}'' Then she decided,  ``\textit{I'll be satisfied with that,}'' and stopped iterating despite the annotation still marked the title as medium level. 
\revi{
The more actionable and \directed annotations seem to enable feedback-driven design iterations, which could bring feelings of progress and creativity, but potentially limiting agency and opportunities for self-experimentation on different design options.}

\paragraph{\Directed Feedback Reduced Design Issues and Did Not Hinder Learning}
\Directed participants improved their design quality by reducing the number of issues in their final designs (see \revise{Table~\ref{tab:quant-summary}}).  Our expert ratings revealed that \directed designs had significantly fewer design principle violations than \baseline designs (\directed: $M= 0.75, SD= 0.72$; \neutral: $ M= 1.54, SD= 0.78$; \baseline: $ M= 1.92, SD= 1.04; U=118, p<.05$). \revi{This aligns with our hypothesis that the most actionable feedback can boost performance.}         

\revised{Contrary to our hypothesis that emphasizing performance might harm learning, we found no significant differences in the principle application learning test across conditions} (\directed: $M= 1.04, SD= 2.16$; \neutral: $ M= 0.73, SD= 1.64$; \baseline: $ M= 0.33, SD= 1.01 $, \revise{non-significant}). 
Our interview data suggest that \revised{\directed feedback did not hinder learning} and 
participants still reflected on how they applied design principles to their designs.
P9 articulated how annotations solidified his understanding of principles by providing specific visual examples: ``\textit{I really think it was (helpful for learning) because reading the design principles, I had an idea for it, and then the visuals, it gave me another example, and the pictures really (show) what it should look like, which added onto the readings which set it in place, OK this is what I should do.}'' Other \directed participants echoed this point. 
P3 initially found it difficult to address design principles, but \textit{\systemname} guided her to apply them. She explained, ``\textit{I needed it to point out where the issues were ... even with the alignment, it looked good to me. But (\systemname) pointed out that it wasn't conceptually correct and then I had to change}'' 
This indicates that \directed annotations \revised{did not hinder learning by virtue of} showing example solutions and prompting participants to apply them within their own design task. 

\paragraph{\Neutral Feedback Encouraged Self-Reflection}
Participants with \neutral feedback regularly reflected on their designs based on annotations and made iterations.
Of all feedback viewed, \neutral participants actively reflected on 34\% of 149 feedback \revised{and} \directed participants reflected on 15\% of 116 feedback during the think-aloud design task. \revise{This aligned with what instructors from the co-design study anticipated: even without exact instruction, students could still understand \neutral annotations since they provided enough information to help novices reflect and make design decisions.}

Indeed, as \neutral participants frequently reflected on the rationale behind their design choices, it led to concrete actions for iterations or reinforced their current decisions. For example, after checking the unity feedback, P22 noticed that she used two different font colors. She then justified that this aligned with her intention since the title should be emphasized with a different color. In another case, when she noticed that two texts had different priority levels from the hierarchy annotation, she updated the font sizes immediately to give title more visual prominence. P22 reflected, ``\textit{It seems like this one is high, this one's medium, but this one should be in the high hierarchy because it is the title of the whole design. So that's why in order to make that the title, I'm going to make the font bigger, maybe 60, super big.}'' 

Participants' reflection also extended beyond the highlighted text elements and led to creative exploration for other design elements, such as images and graphics. From the whitespace annotation, many participants noticed the large empty space on the canvas, which inspired them to add extra elements for a more appealing and balanced design.
As P15 mentioned, ``\textit{there's still a lot of white space on the bottom, so maybe I can potentially add an icon somewhere that I can find in the graphics.}'' However, none of the \directed participants realized such empty space since they were focused on the identified issues and overlooked other visual cues and information communicated by the annotation.

\section{Discussion}\label{sec:discussion}
\revi{We co-designed with experts a set of visual annotations across the feedback actionability spectrum and embedded them into \textit{VizCrit}, an interactive visual design tool supported by algorithmic issue detection and annotation generation. With \textit{VizCrit}, we studied the impact of feedback actionability with design novices and found that \directed feedback improved design quality but may have inflated novices' self-perceived creativity. Some novices requested even more actionability than \directed feedback, suggesting a preference for performance over learning.} 
We review the results and contextualize the limitations and potential future work to extend the exploration around feedback actionability. We then discuss how to calibrate feedback actionability to balance productivity and creative growth, roles of AI in Human-AI co-creativity, and end with 
broader ethical considerations of using AI in creativity support tools. 

\subsection{Limitations and Future Work}


We contextualize our study on feedback actionability for design novices by discussing its limitations and future work.
\subsubsection{Impacts of Design Expertise on Desire for Actionability} 
Our study was conducted with novice designers who are university undergraduate students with little to no prior design experience. Therefore, our findings on how different levels of feedback actionability affect participants' learning, creativity, and design outcomes are situated within this specific novice population. 
\revise{We believe that it is important to conduct further studies with other populations (e.g., novices with an interest in design, intermediate designers with some art training, novices with less familiarity with AI tools) to discover additional insights.} For instance, \revise{novice designers' self-interest in learning visual design could create a difference in their perceptions of feedback. Future studies could target those who are more motivated to develop design skills in settings like design workshops. Additionally, current college students extensively use AI tools to ``get answers'' in various learning contexts \cite{zastudil2023aiedu, adnin2025undisclosed, Amoozadeh2024trustAI, barrett2023aiwrite}, which might be the reason for prioritizing productivity over reflective thinking.
Future work could further investigate populations who use AI tools less frequently or have less efficiency-oriented motivations for using AI (e.g., getting explanations on learning materials).} Other than design novices, more experienced designers who have more design knowledge could perceive and engage differently with \neutraldirected feedback. Future studies should explore how the level of expertise shapes the interaction strategies and interpretations of critiques \cite{Yuan2016AlmostAE} and whether \neutral design annotations better align with professional workflows with more creative opportunities and freedom. 

\subsubsection{Task Influence on User Motivation.} During the study, we provided participants with an initial seed design with design issues.
Although the current setup exposed them to a full set of design annotations and set a starting stage for novices, it could limit their sense of creative ownership. 
The seed design might also bias outcomes as \directed feedback could appear more effective when issues were obvious and inevitable.
\revise{First,} future studies could vary the design task by asking participants to start from scratch \revise{(ideally on a design task that they are self-motivated to complete)} and investigate \revise{how novices interact with \neutral feedback and \textit{emerging} \directed feedback that identifies issues arising from their own design choices}. \revise{Additionally,} researchers should explore an alternative study setup that includes a comprehensive \textit{\systemname} version where both \neutraldirected annotations are available. 
This could allow researchers to observe whether \revise{design phases,} individual preferences, or design habits affect their choices regarding feedback actionability. 

\subsubsection{\revise{Longer Term Studies of Learning.}} 
\revised{We did not find statistically significant differences in the learning tests across conditions.} 
Our qualitative results suggest that \directed feedback \revised{did not hinder} learning by 
\revised{the virtue of offering novices} concrete experiences of solving design issues \cite{kolb1984experientiallearning, dewey1938learnbydoing, schon1983reflective} \revise{and worked examples \cite{chen2023effect} of designs with no issues}. However, we recognize that learning was hard to measure within a short timeframe, since newly acquired knowledge might not be immediately captured by the evaluation \cite{bloom1956taxonomy} and we anticipate the benefits of \neutral annotations to 
occur over a longer period of time. This is because participants' regular self-reflection could support deeper understanding of design principles and \neutral design annotations left more space for learning by iterating and exploring different design possibilities ~\cite{mcdonald2022uncertain, lousberg2020reflection}. 
Longitudinal studies could therefore better capture the learning trajectories and long-term influence of feedback actionability.

\subsubsection{\revise{Moving Beyond Design Principles.}} 
\textit{\systemname} was primarily designed to visualize explicit \revise{knowledge of design principles}. However, these principles could be further extended to other design concepts (e.g., color contrast) \revise{and even tacit design knowledge \cite{son2024tacit}. 
For example, future systems could incorporate design experts' gaze patterns as an overlay or animation. For \neutral annotations, experts' attention patterns or gaze on focal design elements could be highlighted on the canvas, and for \directed annotations, the experts' potential next actions could be further visualized.}

\subsection{Calibrating Feedback Actionability to Balance Productivity and Creative Growth}
\revi{Our results showcased evidence of a range of the influence scaffolding can have on novices' design processes---\neutral feedback encouraged participants to actively reflect on their design choices, while \directed feedback led to better design quality. 
On one hand, the more actionable scaffolding supported novices by making them feel empowered to consider principles and to explore~\cite{kulkarni2013early,mose2014decisive}. On the other hand, we saw how it structured and limited the breadth of their exploration~\cite{li2023beyond,ryhs2025productiveCSI}.
While more actionability may support performance and productivity, 
it has the greatest risk of overreliance, removing the need for deeper self-reflection and learning~\cite{buccinca2021trust, dewey1938learnbydoing, kobiella2024accomplishment}. } This can be especially harmful if novices prioritize a productivity-focused mindset valuing efficiency over reflective thinking.
Feedback with less actionability, on the other hand, requires designers to engage in self-reflection~\cite{kulkarni2013early,mose2014decisive} and experimentation with their own design language rather than being steered prematurely toward normative solutions~\cite{li2023beyond,ryhs2025productiveCSI}. 

Therefore, we suggest design opportunities for more dynamic computational design tools
\revi{by calibrating feedback actionability 
along several dimensions, including interaction timing, user goals, and experience levels. 
In terms of timing, designers might consider creating tools to progressively increase actionability throughout a design process.} Early on, novices can use less actionable feedback to support ongoing reflection and exploration of the design space, and later, use more actionable feedback to resolve any remaining issues and make refinements. This potentially requires future AI tools to actively identify design phases. 
\revi{To support diverse user goals (e.g., doing a task for the creative experience or just for a quick outcome), users should be able to tailor actionability~\cite{e2024timing,ryhs2025productiveCSI, chan2022investigating}. For example, when they are under time constraints \cite{cao2023time, swaroop2024pressure} and desire efficiency, most actionable feedback can support them in  boosting productivity (potentially accompanied by multiple outcomes in parallel for designers to choose from~\cite{dow2010parallel}). Additionally, users with more experience or who want to prioritize their learning or agency may prefer the flexibility of less actionable feedback, while novices may need more actionability when decision-making still feels intimidating.} 

\subsection{Understanding the Roles of AI in Human-AI Co-Creativity}
By computationally inspecting the design,
\textit{\systemname} not only acts as a collaborator to deliver constructive feedback, but also as an audience to simulate how others perceive the design~\cite{hung2025simtube, benharrak2024writer}. Both \neutraldirected participants described designing ``in others' eyes'' and become more aware of how a professional, artistic reviewer may view and evaluate their draft~\cite{goodwin1994professional, elliott1993vision}, also aligning with prior insights on a tendency to conform too closely to the audience perspectives~\cite{hung2025simtube}.  
Regardless, this can be seen as a form of external validation to users, especially for novices or inexperienced creators, who typically lack confidence in their creative outcomes \cite{Tanner2019recognize} or lack access to timely feedback \cite{dannels2008critiquing}.
This may help encourage novices to adventure a new domain as social apprehension is reduced \cite{krishna2021ready, kotturi2019wild} and experiment with more creative possibilities and seek new feedback after each attempt \cite{Intasao2018BeliefsAC}. 

Beyond individual creative experience, AI tools can also be a social mediator that facilitates collaboration while having a cordial atmosphere \cite{e2020adaptivephotocomposition, suh2021glue}. 
As creativity is highly subjective and personal \cite{simonton2000creativity}, with AI's objective analysis, collaborators can have a shared reference for discussion and psychologically safer spaces for negotiation and creative alignment.
AI tools can further enable collaborators to build and customize creative and learning experiences \cite{li2021Artists}. 
\revise{For example, researchers can investigate how design instructors and professionals adopt or customize \textit{VizCrit}'s feedback in teaching contexts and what additional scaffolds may be needed to support their uses.}

\subsection{Navigating Ethical Considerations of AI in Creativity Support Tools}
Our results highlighted ethical considerations of AI in CSTs regarding its power and the normativeness of scaffolding based on the well-known guidelines, especially given novices' tendency to feel the compliance pressure from rule-based feedback, and their  tendency to overly rely and trust the tool.

\subsubsection{Compliance Pressure from Rules-based Feedback}
Novices often perceive design principles as strict rules to follow and memorize instead of flexible guidelines to adapt or break ~\cite{abbott2014system}. As novices tended to lack confidence in breaking preset rules and inappropriately relied on ``AI'' solutions \cite{ma2024selfconfidence}, they could easily interpret \directed feedback as perfect answers that delineated the correct and problematic attempts. As a result, novices might be compelled to make changes to satisfy the system, giving it more power over their creative decision-making process~\cite{li2023beyond}. In contrast, \neutral feedback allowed more exploration within the constraints of design principles. Without knowing a system-side definite answer, novices had spaces 
to reflect, justify, and reinforce their design choices, even though they might be  ``wrong'' based on principles. Therefore, to better maintain creative agency, CSTs with AI could aim to empower novices to decide when and how to follow feedback and provide breathing room to question computational analysis~\cite{dow2010parallel,terry2004variation}.

\subsubsection{Overreliance on AI and False Sense of Creativity}
The contrasting result from \directed participants' self-perceived and expert-rated creativity uncovered their false sense of creativity and unconscious overreliance on the system.
Novices can easily overly trust AI \cite{buccinca2021trust, benda2021trust} to evaluate and revise their designs, which not only risks agency, but leads to even worse outcomes if the system provides feedback that follows principles (algorithmically correct), but is not applicable in a personal creative context. 
Prior work has shown that explanations can help reduce overreliance and facilitate creative outcomes ~\cite{vasconcelos2023explanations, Kulesza2013explanation, ahmed2019dyads}. However, in our case we thought the annotations would serve as easier-to-understand \textit{explanations} (than traditional text feedback), but it instead might have further created a false sense of trust.
This implies that AI should encourage more critical self-reflection to avoid the possibility of being a substitution of creative decision-making. Computational tools could not only help novices with efficiency gains, but also skill building and creative growth for better retaining user agency and avoiding overreliance.

\section{Conclusion}\label{sec:conclusion}
As AI systems make instant, on-demand design feedback widely available, it is essential to understand how feedback actionability shapes creative work. Informed by co-design interviews with nine experts, we introduce \textit{VizCrit}, a prototype visual design tool that automatically detects design issues and generates multi-modal feedback, including visual annotations, across the actionability spectrum: \baseline, \neutral, and \directed.
In a between-subjects study with 36 novices, \revised{we discover that \directed feedback supported design quality, did not hinder learning, and increased self-perceived creativity. However, this self-perception of creativity may be inflated by virtue of feedback-driven iterations and sense of progress. These results argue that less actionable, \neutral feedback encourages self-reflection, while more actionable, \directed feedback supports performance and productivity but may risk overreliance.}
For designers of AI-powered creativity support tools, we recommend calibrating feedback  actionability for design novices \revi{considering the interaction timing and users' goals.} In doing so, systems can pair immediate assistance with long-term growth in judgment, reflection, and creative exploration.

\begin{acks}
We thank our participants and reviewers for their time and constructive feedback.
We also thank Isabelle Pan, Grace Lin, Yu-Chun Grace Yen, Hyoungwook Jin, Fuling Sun, Chenrong Gu, and Teguh Hofstee for their support, including initial explorations, pilot testing, and discussion around project ideas.
We appreciate the financial support from UC San Diego's CSE department, with special thanks to Sorin Lerner, the Stanford HAI
Postdoctoral Fellowship, and the National Science Foundation (\#2009003).
\end{acks}

\bibliographystyle{ACM-Reference-Format}
\bibliography{realtime-annotations}

\appendix
\section{Appendix}\label{sec:appendix}

\subsection{Design Probe Co-Design}
\begin{figure*}[th]
    \centering
    \includegraphics[width=\linewidth]{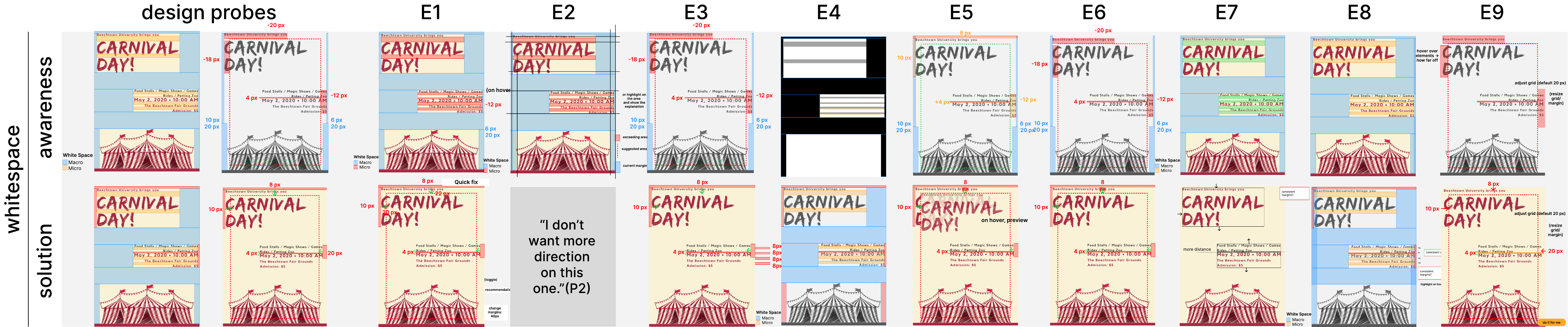}
    \caption{Iterative design of the whitespace design probes: our set of 4 annotations at the left for whitespace (2 \neutral top, 2 \directed bottom), along with the expert iterations of each. Most started with their preferred of our two designs and made revisions from there. 
    }
    \Description{
    This figure shows the early iterative design process with instructors. It shows two rows of designs, the top all awareness-centered and the bottom all solution-centered. Left to right, it shows our design probes that color-code larger and smaller whitespace in two different colors, sometimes also listing pixel values of the distances. After our design probes are all the participants' designs mostly modified off of ours. Two that are most different had one block out all text in white and everything else in black for awareness-centered (E4), and the other said “I don't want more direction on this one” for solution-centered (E2).
    }
    \label{fig:designprobes_whitespace}
\end{figure*}

This was an example of design probe annotations for the whitespace principle. Co-design participants iterated on their preferred design for both \neutraldirected annotations (Figure \ref{fig:designprobes_whitespace}).

\subsection{Annotation Explanations}
Each annotation had corresponding text explanations in table form. These are the range of possible explanations in the system across \neutral and \directed annotations (Table
\ref{tab:tabletexts}).

\begin{table*}[ht]
\centering
\caption{Explanations for visual annotations: annotations on the canvas will have explanations in the text format on the design principles panel. The content in the square bracket indicates the specific fields that will appear in the text explanation table.}
\begin{tabular}{|
  p{2cm}|
  p{7.5cm}|
  p{7cm}|
}
\hline
\textbf{Principle} & \textbf{\Neutral Explanations} & \textbf{\Directed Explanations} \\
\hline

\textbf{Hierarchy} &
[Text element] in the design has [level of emphasis] visual emphasis. &
The [title] does not seem particularly emphasized compared to [competing text element(s)], making it difficult to recognize what is most important.\par
The [title] does not seem particularly emphasized. \\
\hline

\textbf{Alignment} &
Text in the design forms x alignment groups with [text element(s)]. [Text element] is [internal alignment]-aligned. &
Text in the design forms many separate alignment groups with [text element], which can make the design appear somewhat incohesive.\par
[Text element(s)]'s relative positioning on the canvas skews [external alignment], which does not match with the text being [internal alignment]-aligned. \\
\hline

\textbf{Whitespace} &
[Text element] is on the [edge] side of the canvas. Note other nearby elements and the lengths of the lines of text.\par
[Text element] and its shape is on the [edge] side of the canvas with other nearby elements. &
The line breaks in [text element] creates uneven text lengths, which can somewhat disrupt the visual flow of the text.\par
[Text element] placed quite close to the [edge], which can make the design appear somewhat crowded.\par
[Text elements] placed quite close to each other, which can make the design appear somewhat crowded. \\
\hline

\textbf{Unity} &
[Text element] in the design has the following text properties: [font family], [style], [size], [color]. &
[Text elements] in the design use many different text properties: [font family], [style], [size], [color], which can make your design seem incohesive. \\
\hline

\end{tabular}
\label{tab:tabletexts}
\end{table*}

\subsection{Principle Details}
These were the descriptions that could be surfaced in the interface when participants clicked on ``show'' to see more about the principle or how to apply the principle (Table \ref{tab:principledetails}).

\begin{table*}[tbp]
\small
\centering
\caption{Principle details: the full principle information after expanding ``Learn More About'' on the design principles panel.}
\begin{tabular}{|p{1.9cm}|p{6cm}|p{8.5cm}|}
\hline
\textbf{Principle} & \textbf{Learn More About Principle} & \textbf{Learn More About Applying This Principle} \\
\hline

\textbf{Hierarchy} &
\textbf{Principle Definition:}\par
Hierarchy is the visual level of emphasis of each element in your design—this should match the importance of the content. A common issue is having a weak point of entry for the viewer.\par
\textbf{Common Issue:}\par
The point of entry is the focal (most eye-catching) point of your design. If the point of entry is weak, it is unclear what element is the most important and viewers may not recognize the key message at first sight. &
\textbf{Example:}\par
On a website, a large, bold headline at the top of the page establishes that it's the most important element, while smaller subheadings and body text are visually less prominent, guiding the reader's eye through the content.\par
\textbf{Related Actions:}\par
Increase importance of title\par
- make title text larger\par
- make title different: e.g., bold, color, font, all caps/lowercase, etc.\par
- use more prominent colors/fonts\par
- use additional white space as frame\par
Decrease importance of non-title text\par
- make body text smaller\par
- reduce brightness of body text\par
- use neutral font for body text \\
\hline

\textbf{Alignment} &
\textbf{Principle Definition:}\par
Alignment refers to the relative positioning of elements in your design. A common issue is for elements to appear to have somewhat arbitrary alignment relative to each other.\par
\textbf{Common Issue:}\par
If several elements aren't aligned to common axes or boundaries, the design may appear to lack a sense of order or organization. &
\textbf{Example:}\par
In a webpage layout, aligning text blocks to the left or centering them on the page creates a structured flow that helps the viewer follow the content easily.\par
\textbf{Related Actions:}\par
Create and align common boundaries (external alignment)\par
- align boundaries of nearby text boxes\par
- align to closest margin (e.g. left page margin)\par
- align to common shape (e.g. image)\par
Match internal and external alignment\par
- left/right align text that is positioned on towards the left/right margins\par
- center position text that is center aligned\par
- consider using the same internal alignment (e.g. left-aligned) for related groups of text \\
\hline

\textbf{Whitespace} &
\textbf{Principle Definition:}\par
Whitespace refers to the spacing between and around elements in your design. A common issue is to have insufficient or ragged spacing relative to other elements or the margins.\par
\textbf{Common Issue:}\par
If a block of text has too little margins relative to other elements or the edge of the canvas, or the text lengths are uneven, it can make the design appear somewhat crowded. &
\textbf{Example:}\par
The empty space around a logo or between paragraphs allows the content to be more digestible, making the design feel less crowded and more inviting.\par
\textbf{Related Actions:}\par
Elements should have breathing room/space\par
- give elements space away from boundaries/margins\par
- leave whitespace between elements\par
- consider size of elements when determining how much whitespace should surround it (bigger elements might need more space)\par
Avoid uneven lengths of text within an element\par
- avoid widows, orphans, and singles in your text layout\par
- consider adjusting text box width to make lines more even \\
\hline

\textbf{Unity} &
\textbf{Principle Definition:}\par
Unity refers to the cohesion in your design. A common issue is to have too many variations in fonts/styles across different elements such as text.\par
\textbf{Common Issue:}\par
Using too many variations in font families, sizes, styles, and colors can make a design look incoherent and unorganized. &
\textbf{Example:}\par
In branded materials, using consistent colors, logo styles, and fonts across all materials (like posters, brochures, and websites) creates a unified look and reinforces the brand identity.\par
\textbf{Related Actions:}\par
Limit number of font families, font sizes, font colors, font styles, etc.\par
- reduce total number of fonts/sizes/colors/styles\par
- make similar fonts/sizes/colors/styles the same\par
- make different fonts/sizes/colors/styles more different\par
Use text styles in a consistent manner\par
- treat text of equal importance with same font/size/color/style \\
\hline

\end{tabular}
\label{tab:principledetails}
\end{table*}

\subsection{System Implementation Details}\label{sec:implement}
We describe the implementation of how the server communicates with the user interface. The frontend interface is built on top of Polotno\footnote{https://polotno.com/}, an open-sourced web-based visual design tool. The server is a Python backend, using Pillow\footnote{https://pillow.readthedocs.io/en/stable/reference/ImageDraw.html} to generate annotation images. A websocket server facilitates communication between them.
The interface needs to provide updated data on the user's design to the server, and the server provides annotations and associated information back to the interface to surface to the user. Note that the tool runs locally in its current form.

\paragraph{Retrieving the Current On-Canvas Design}
To analyze a user design in real-time, the interface monitors on-canvas design activities. 
Whenever there is a change on the canvas (with 4-sec debounce time to avoid over-requests),
the interface sends the current design data to the server, including a canvas JSON object, full design image, and text-only image (for more close analysis of space occupied by text). Images were generated using Polotno Cloud Rendering API\footnote{https://polotno.com/cloud-render}. Algorithms primarily used the JSON description of the canvas, with a few requiring some level of image understanding directly from the full design or text-only images.

\paragraph{Generating Annotation-related Data} Once the server receives the design data, it runs algorithms to generate annotations across all principles. Updated annotations are saved in a location that is accessible by the interface and all other data is sent back via JSON to be used to populate the interface.



\end{document}